\documentclass[aps,prd,showpacs,amsmath,10pt,twocolumn,superscriptaddress,floatfix,nofootinbib,notitlepage]{revtex4-1}

\usepackage[utf8]{inputenc}
\usepackage[english]{babel}
\usepackage{bm,color,xcolor,amsfonts,amsmath,amssymb,epsfig,esint}

\usepackage[normalem]{ulem}

\newcommand{\be}{\begin{equation}}
\newcommand{\ee}{\end{equation}}
\newcommand{\bea}{\begin{eqnarray}}
\newcommand{\eea}{\end{eqnarray}}
\newcommand{\beas}{\begin{eqnarray*}}
\newcommand{\eeas}{\end{eqnarray*}}
\newcommand{\ds}{\displaystyle}

\newcommand{\Br}{\mbox{Br}}

\newcommand{\zb}{Z_b^{\pm}(10610)}
\newcommand{\zbp}{Z_b^{\pm}(10650)}
\newcommand{\zc}{Z_{c}^{\pm}(3900)}
\newcommand{\zcp}{Z_{c}^{\pm}(4020)}
\def\vec#1{\boldsymbol{#1}}

\newcommand{\bra}{\langle}
\newcommand{\ket}{\rangle}
\newcommand{\Omnes}{Omn\`{e}s }

\newcommand{\bonn}{\affiliation{Helmholtz-Institut f\"ur Strahlen- und Kernphysik and Bethe Center for Theoretical Physics,\\ Universit\"at Bonn, D-53115 Bonn, Germany}}

\newcommand{\fzj}{\affiliation{Institute for Advanced Simulation, Institut f\"ur Kernphysik and J\"ulich Center for Hadron Physics, Forschungszentrum J\"ulich, D-52425 J\"ulich, Germany}}

\newcommand{\itep}{\affiliation{Institute for Theoretical and Experimental Physics NRC ``Kurchatov Institute'', Moscow 117218, Russia }}

\newcommand{\lebedev}{\affiliation{P.N. Lebedev Physical Institute of the Russian Academy of Sciences, 119991, Leninskiy Prospect 53, Moscow, Russia}}

\newcommand{\mipt}{\affiliation{Moscow Institute of Physics and Technology, 141700, Institutsky lane 9, Dolgoprudny, Moscow Region, Russia}}

\newcommand{\rub}{\affiliation{Institut f\"ur Theoretische Physik II, Ruhr-Universit\"at Bochum, D-44780 Bochum, Germany}}

\graphicspath{{figs.dir/}}

\begin{document}

\title{Insights into $Z_b(10610)$ and $Z_b(10650)$ from dipion transitions from $\Upsilon(10860)$}

\author{V.~Baru}
\bonn \itep \lebedev

\author{E.~Epelbaum}
\rub

\author{A.A.~Filin}
\rub

\author{C.~Hanhart}
\fzj

\author{R.V.~Mizuk}
\lebedev

\author{A.V.~Nefediev}
\lebedev \mipt

\author{S.~Ropertz}
\bonn

\begin{abstract}
The dipion transitions $\Upsilon(10860)\to\pi^+\pi^-\Upsilon(nS)$ ($n=1,2,3$) are studied in the framework of a unitary and 
analytic coupled-channel formalism previously developed for analysing experimental data on the bottomoniumlike states $Z_b(10610)$ and $Z_b(10650)$ [Phys. Rev. D {\bf 98}, 074023 (2018)] and predicting the properties of their spin partners [Phys. Rev. D {\bf 99}, 094013 (2019)]. In this work we use a relatively simple but realistic version of this approach, where the scattering and production amplitudes are constructed employing only short-ranged interactions between the open- and hidden-flavour channels consistent with the constraints from heavy quark spin symmetry, for an extended analysis of the experimental line shapes. In particular, the transitions from the $\Upsilon(10860)$ to the final states $\pi \pi h_b(mP)$ ($m=1,2$) and $\pi B^{(*)}\bar B^* $
already studied before, are now augmented by the $\Upsilon(10860)\to\pi^+\pi^-\Upsilon(nS)$ final states ($n=1,2,3$). This is achieved by employing dispersion theory to account for the final state interaction of the $\pi\pi$ subsystem including its coupling to the $K\bar K$ channel. Fits to the two-dimensional Dalitz plots for the $\pi^+\pi^-\Upsilon$ final states were performed. 
Two real subtraction constants are adjusted to achieve the best description of the Dalitz plot for each $\Upsilon(nS)$ ($n=1,2,3$) while all the parameters related to the properties of the $Z_b$'s are kept fixed from the previous study. A good overall description of the data for all $\Upsilon(10860)\to\pi^+\pi^-\Upsilon(nS)$ channels achieved in this work provides additional strong support for the
molecular interpretation of the $Z_b$ states. 
\end{abstract}



\maketitle


\section{Introduction}

The spectroscopy of hadronic states containing heavy quarks remains one of the fastest developing and most intriguing branches of strong interaction studies. Many new states have been discovered in the spectrum of charmonium and bottomonium which do not fit into the quark model scheme and qualify as exotic states. For example, the states 
$\zb$, $\zbp$ \cite{Belle:2011aa}, $\zc$ \cite{Ablikim:2013mio,Liu:2013dau}, $\zcp$ \cite{Ablikim:2013wzq}, $Z^\pm(4430)$~\cite{Choi:2007wga,Mizuk:2009da,Chilikin:2013tch,Aaij:2014jqa} 
are charged and decay into final states containing a heavy quark and its antiquark. Since the production of this pair of heavy quarks in the decay is highly suppressed in QCD, it must have been present in the wave functions of the states and as such the $Z_Q$ states cannot be conventional $\bar{Q}Q$ (with $Q$ denoting a heavy quark) mesons as they must contain at least four quarks. The interested reader can find a comprehensive overview of the current experimental and theoretical status of the exotic hadrons with heavy quarks in dedicated review papers, for example, in Refs.~\cite{Lebed:2016hpi,Esposito:2016noz,Ali:2017jda,Guo:2017jvc,Olsen:2017bmm,Liu:2019zoy,Brambilla:2019esw}. 

The $\zb$ and $\zbp$ bottomoniumlike states (in what follows often referred to as $Z_b$ and $Z_b'$, respectively) are ideally suited for both experimental and theoretical studies since there exist two resonances in the same $J^{PC}=1^{+-}$ channel \cite{Collaboration:2011gja} 
split by only about 40 MeV which are simultaneously
seen in several modes. Specifically, the Belle Collaboration observed them as distinct peaks (i) in the invariant mass distributions of the $\pi^\pm\Upsilon(nS)$ ($n=1,2,3$) and $\pi^\pm h_b(mP)$ ($m=1,2$) subsystems in dipion transitions from the vector bottomonium $\Upsilon(10860)$ \cite{Belle:2011aa} and (ii) in the elastic 
$B\bar{B}^{*}$\footnote{A properly normalised $C$-odd combination of the $B\bar{B}^*$ and $\bar{B}B^*$ components is understood.} 
and $B^{*}\bar{B}^{*}$ 
channels in the decays $\Upsilon(10860)\to\pi B^{(*)}\bar{B}^*$ \cite{Adachi:2012cx,Garmash:2015rfd}. 
The two most prominent explanations for the $Z_b$'s claimed to be consistent with the data are provided by a tetraquark model \cite{Ali:2011ug,Esposito:2014rxa,Maiani:2017kyi} and a hadronic molecule picture 
\cite{Bondar:2011ev,Cleven:2011gp,Nieves:2011vw,Zhang:2011jja,Yang:2011rp,Sun:2011uh,Ohkoda:2011vj,Li:2012wf,Ke:2012gm,Dias:2014pva}. A review of the sum rules approach to the exotic states with heavy quarks and relevant references on the subject can be found in a recent review \cite{Albuquerque:2018jkn}. It should be noted that a particularly close location of the $Z_b$'s to the thresholds of the $B\bar{B}^*$ and $B^*\bar{B}^*$ channels, which in addition are the most dominant decay modes for them, provides a strong hint in favour of their molecular interpretation. 

Both the $\zb$ and $\zbp$ contain a heavy $b\bar{b}$ pair, so it is commonly accepted that the 
heavy-quark spin symmetry (HQSS) should 
be realised to high accuracy in these systems and indeed, HQSS is able to explain naturally the interference pattern in the inelastic channels $Z_b^{(\prime)}\to\pi\Upsilon(nS)$ and $Z_b^{(\prime)}\to\pi h_b(mP)$ \cite{Bondar:2011ev}.
In Ref.~\cite{Wang:2018jlv}, an effective field theory (EFT) approach to the $Z_b$ states consistent with HQSS and chiral symmetry was developed to perform a combined analysis of the experimental data in the channels
\bea
\Upsilon(10860)&\to&\pi Z_b^{(\prime)}\to \pi B^{(*)}\bar{B}^*,\nonumber\\[-2mm]
\label{chains}\\[-2mm]
\Upsilon(10860)&\to&\pi Z_b^{(\prime)}\to\pi\pi h_b(mP),\quad m=1,2.\nonumber
\eea

The information on the branching fractions in the transitions $\Upsilon(10860)\to\pi Z_b^{(\prime)}\to\pi\pi\Upsilon(nS)$ ($n=1,2,3$) was also used, but no analysis of the line shapes in these channels was performed for the reasons explained below. 
A fairly good description of the data was achieved in different fitting schemes described in detail in Ref.~\cite{Wang:2018jlv}. As expected, the experimental data on the $Z_b$'s are fully consistent with HQSS, since symmetry violating terms in the effective hadronic potential are argued to play a minor role~\cite{Wang:2018jlv}. 
In Ref.~\cite{Baru:2019xnh}, the approach was extended to predict in a parameter-free way the properties of the spin partner states
of the $Z_b$'s, the $W_{bJ}$'s ($J=0,1,2$).

The one-pion exchange (OPE) in the bottomoniumlike systems under consideration was a special concern of the quoted works \cite{Wang:2018jlv,Baru:2019xnh}, and it was concluded to play an important role for the 
$Z_b$'s and $W_{bJ}$'s. Indeed, the poles of the $Z_b$’s and $W_{bJ}$'s that were originally classified as virtual states in the pionless framework 
moved above the nearby elastic thresholds to become resonances, as an effect of the OPE. 
Meanwhile, the conclusion that all these states are hadronic molecules, based on
a decent description of the data, follows already from the scheme with purely contact interactions in the $B^{(*)}\bar{B}^*$ system 
(Scheme A, in the notation of Ref.~\cite{Wang:2018jlv}, yields $\chi^2/N_{\rm dof}\approx 1.23$). Note also that this fitting 
scheme provides results identical to those obtained with the help of an analytical parametrisation for the line shapes derived previously in Refs.~\cite{Hanhart:2015cua,Guo:2016bjq}. 

Not all experimental information used in the aforementioned combined analysis could be considered on equal footing. Indeed, while the line shapes in the $\pi h_b(mP)$ and $B^{(*)}\bar{B}^*$ channels could be fitted directly, 
as discussed above, only the total branchings for the $\pi\Upsilon$ final states were used in the fit. The signal in the latter channels contains a significant nonresonant contribution that depends on the invariant mass of the two-pion system, so that the amplitude analysis has to be multidimensional. This analysis is in the spotlight of the present work. 
In particular, we generalise the approach developed in Ref.~\cite{Wang:2018jlv} to incorporate coupled-channel effects from the $\pi\pi$-$K\bar K$ interactions in the final state using a model-independent dispersive approach. Then we 
perform maximum likelihood fits to the Dalitz plots of the reactions $\Upsilon(10860)\to\pi\pi\Upsilon(nS)$ ($n=1,2,3$). 
To keep consistency with the data in the $\pi h_b(mP)$ and $B^{(*)}\bar{B}^*$ channels, we directly employ the inelastic production amplitudes
obtained in Ref.~\cite{Wang:2018jlv} 
for Scheme A as input for the present research. 
Since the focus of the present study is on the development of the dispersive treatment of the final state interactions (FSI),
we resort to a simple pionless formulation, as provided by Scheme A, while effects from the OPE will be included in future studies.
Thus in this study we focus on the following goals: 
\begin{itemize} 
\item [(a)] A development of a dispersive approach to the $\Upsilon(10860)\to\pi\pi\Upsilon(nS)$ transitions and a systematic account for the effects from the $\pi\pi$ FSI including the coupling to the $K\bar K$ channel. While for the $\Upsilon(2S)$ and, especially, $\Upsilon(3S)$ in the final state the $\pi\pi$-$K\bar K$ coupling is expected to play a marginal role, it should be important for the $\Upsilon(1S)$ channel near the $K\bar K$ threshold (see, for example, Ref.~\cite{Surovtsev:2015hna}). This effect can be included in a 
model-independent way using an \Omnes matrix
constructed from high accuracy determinations of the $\pi\pi$ and $K\bar K$ scattering amplitudes as well 
as from the $B_s$ decay data~\cite{Daub:2015xja,Ropertz:2018stk}.

\item[(b)] Our focus is on the inclusion of the FSI while keeping the full complexity of the $Z_b$ dynamics, so we consider two production mechanisms
for the transitions $\Upsilon(10860)\to\pi\pi\Upsilon(nS)$, namely (i) through the contact operators with two real parameters and (ii) through $B$-meson production assuming pointlike vertices with the subsequent $B$-meson 
interactions in the final state, that is,
via the process $\Upsilon(10860)\to B^{(*)}\bar B^{*}\pi\to \pi\pi\Upsilon(nS)$. Both mechanisms are supplemented with the $\pi\pi$ FSI. 
Note that in Ref.~\cite{Chen:2016mjn} also a possible impact of the box-diagram mechanism was studied, which is not included here. 
The underlying rationale is that in the $\Upsilon(10860)$ decays the $Z_b$ states can go on-shell and should
by far dominate the effects from the $B^{(*)}B^*$ intermediate states.
The corresponding imaginary parts in the production amplitudes
 are taken into account explicitly in this work. As a consequence, only two real subtraction constants defined in the mechanism (i) are sufficient to dispersively reconstruct the amplitude, which is insensitive to the high-energy 
integration range. 
This is unlike to Ref.~\cite{Molnar:2019uos}, where two complex coefficients were utilized in a related study of the dipion transitions in the charmonium sector. 
\item[(c)] The Dalitz plots for the $\Upsilon(10860)\to\pi\pi\Upsilon(nS)$ transitions contain nontrivial information about the $Z_b$'s --- these states can be clearly seen in the $\pi\Upsilon(nS)$ invariant mass distributions and have imprint also on the $\pi\pi$ spectrum. Thus we analyse the two-dimensional Dalitz plots to check whether the results for the $Z_b$'s from our previous analyses are consistent with them. 
\end{itemize} 

The paper is organised as follows. In Sect.~\ref{sec:coupled} we briefly introduce the coupled-channel approach suggested and used in Refs.~\cite{Wang:2018jlv,Baru:2019xnh}. In Sect.~\ref{sec:fsi} a dispersive approach to the decay amplitude is developed to take into account the $\pi\pi$ interaction in the final state. Section \ref{sec:analysis} is devoted to the data analysis for the reactions $\Upsilon(10860)\to\pi\pi\Upsilon(nS)$ ($n=1,2,3$). Our conclusions are discussed in Sect.~\ref{sec:conslusions}. Appendices~\ref{app:Omega} and \ref{app:anom} provide some technical details of the dispersive approach used in this work, including a discussion of the anomalous contributions to the amplitude.

\section{Coupled-channel approach}
\label{sec:coupled}
\begin{table*} 
 \begin{tabular}{|c|c|c|c|c|c|c|c|}
 \hline 
 Parameter & $\mathcal{C}_{d}$, GeV$^{-2}$& $\mathcal{C}_{f}$, GeV$^{-2}$& $|g_{\Upsilon(1S)}|$, GeV$^{-2}$ & $|g_{\Upsilon(2S)}|$, GeV$^{-2}$ & $|g_{\Upsilon(3S)}|$, GeV$^{-2}$ & $|g_{h_{b}(1P)}|$, GeV$^{-3}$ & $|g_{h_{b}(2P)}|$, GeV$^{-3}$ \tabularnewline
 \hline 
 \hline 
 Value &$-3.30\pm0.11$& $-0.06\pm0.13$ &$0.30\pm0.07$ & $1.01\pm0.20$ & $1.28\pm0.34$ & $3.29\pm0.38$ & $11.38\pm 1.46$ \tabularnewline
 \hline
 \end{tabular}
 \caption{\label{tab:par}The fitted values of the contact terms and inelastic coupling constants for Scheme A from Ref.~\cite{Wang:2018jlv}. The cut-off $\Lambda$ is set to 1~GeV (see the discussion in the quoted paper). For the inelastic coupling constants only the absolute values are presented since physical quantities are not sensitive to their signs.} 
 \end{table*}
 
 \begin{figure*}[t]
 \begin{center}
 \epsfig{file=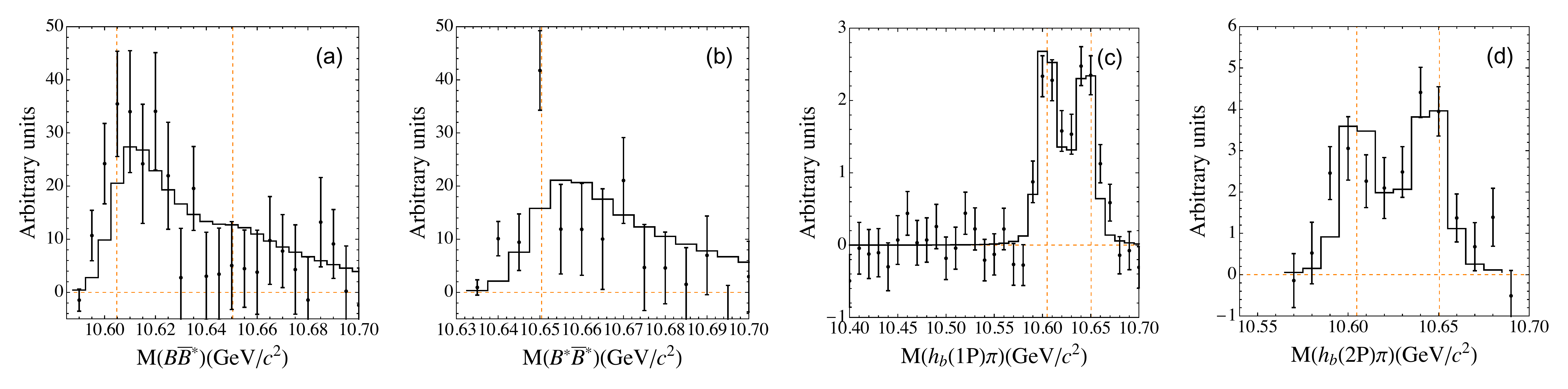,width=0.99\textwidth}
 \end{center}
 \caption{The line shapes in the elastic channels $B\bar{B}^*$ and $B^*\bar{B}^*$ and inelastic channels $\pi h_b(mP)$ ($m=1,2$) provided by Scheme A from Ref.~\cite{Wang:2018jlv}. Experimental data from Refs.~\cite{
 Belle:2011aa,Garmash:2015rfd} are shown as dots with error bars.}\label{fig:fita}
 \end{figure*} 

In this section we briefly recall some essentials of the coupled-channel approach previously developed in Ref.~\cite{Wang:2018jlv} to perform a combined analysis of the data for 
the bottomoniumlike states $Z_b(10610)$ and $Z_b(10650)$. The channels with hidden bottom (labelled by latin letters),
\be
i=\pi\Upsilon(1S),~\pi\Upsilon(2S),~\pi\Upsilon(3S),~\pi h_b(1P),~\pi h_b(2P),
\label{inelch}
\ee
are referred to as inelastic ones while the open-bottom channels (labelled by greek letters),
\be
\alpha=B\bar{B}^*,~ B^*\bar{B}^*,
\label{elch}
\ee 
are denoted as elastic ones. The interaction potential between different channels takes the form of a matrix,
\begin{equation}
 V^{\rm pionless}=
 \begin{pmatrix}
 v_{\alpha\beta}(p,p') & v_{\alpha i}(p,k_i)\\
 v_{j\beta}(k'_j,p') & v_{ji}(k'_j,k_i)
 \end{pmatrix}.
 \label{pot}
 \end{equation}

The main purpose of the present work is to incorporate into the current coupled-channel scheme of Ref.~\cite{Wang:2018jlv} 
 the pion interaction in the $\pi\pi\Upsilon(nS)$ final states.
Consequently, although the OPE was argued in Ref.~\cite{Wang:2018jlv} to provide an important contribution to the elastic potential, its analytic structure is quite complicated and had the largest impact on the $B\bar B^*$ channel.
Accordingly, it will be neglected in the current study, especially since
 existing data can be quite well described within the purely contact Scheme A of Ref.~\cite{Wang:2018jlv}. Thus, in what follows, we stick to this scheme and assume that the leading left-hand cut contributions to the amplitude
 for the $\pi\pi$ FSI are generated by the $Z_b$'s poles and the $B^{(*)}\bar{B}^{*}$ cuts which are taken into account in the present approach.
To be specific, in this work we employ the following approximations: 
\begin{itemize}
\item Only $O(p^0)$ contact interactions are included in the elastic channels, so that the $v_{\alpha\beta}(p,p')$ from Eq.~(\ref{pot}) takes the form
\begin{eqnarray}
 v\left(p,p^\prime \right)=\left(\begin{array}{cc}
 \mathcal{C}_{d}& \mathcal{C}_{f}\\
 \mathcal{C}_{f}& \mathcal{C}_{d}\\
 \end{array}\right),
 \label{vSS}
 \end{eqnarray}
where $\mathcal{C}_{d}$ and $\mathcal{C}_{f}$ are independent low-energy constants;
\item Elastic-to-inelastic transition potentials are parametrised via coupling constants as
\be
v_{i\alpha}(k_i,p)=v_{\alpha i}(p,k_i)=g_{i\alpha} k_i^{l_i}, 
\label{via}
\ee
where $k_i$ and $l_i$ are the momentum and the angular momentum in the $i$-th inelastic channel, respectively. 
The inelastic momentum is calculated as 
\be
k_i=\frac{1}{2M}\lambda^{1/2}(M^2,m_{H_i}^2,m_{h_i}^2),
\label{eq:mominelastic}
\ee
where $m_{H_i}$($m_{h_i}$) is the mass of the heavy(light) meson in this channel, $M$ is the total energy of the system, and $\lambda(m_1^2,m_2^2,m_3^2)$ is the
standard K\"allen triangle function,
\be
\lambda(x,y,z)=x^2+y^2+z^2-2xy-2xz-2yz.
\label{lambda}
\ee

The coupling constants $g_{i\alpha}$ are constrained by HQSS:
\bea
&\ds \frac{g_{[\pi\Upsilon(nS)][B^*\bar{B}^*]}}{g_{[\pi\Upsilon(nS)][B\bar{B}^*]}}=-1,\quad n=1,2,3,&\nonumber\\[-2mm]
\label{HSconstr}\\[-2mm]
&\ds \frac{g_{[\pi h_b(mP)][B^*\bar{B}^*]}}{g_{[\pi h_b(mP)][B\bar{B}^*]}}=1,\quad m=1,2.&\nonumber
\eea
Therefore, as in Ref.~\cite{Wang:2018jlv}, they will be quoted only for the $B\bar{B}^*$ channel in the form
\bea
&g_{\Upsilon(nS)} \equiv g_{[\pi\Upsilon(nS)][B\bar{B}^*]},&\nonumber\\[-2mm]
\\[-2mm]
&g_{h_b(mP)} \equiv g_{[\pi h_b(mP)][B\bar{B}^*]}.&\nonumber
\eea
\item Following the arguments from Refs.~\cite{Hanhart:2015cua,Guo:2016bjq}, direct interactions in the inelastic channels are neglected, $v_{ji}(k_j',k_i)=0$.
\item The long-ranged part of the pion exchange between the $B^{(*)}$ mesons is not considered\footnote{As was demonstrated in Ref.~\cite{Wang:2018jlv}, the short-range central part of the OPE can be absorbed effectively 
into the low-energy constants $\mathcal{C}_d$ and $\mathcal{C}_f$.}.
\end{itemize} 

As a result of these approximations, the effective elastic-to-elastic channel transition potential takes the form
\be
V^{\text{eff}}_{\alpha\beta}(M,p,p')=v_{\alpha\beta}(p,p')-i\sum_i\frac{m_{H_i} m_{h_i} }{2\pi M} g_{i\alpha} g_{i\beta}k_i^{2 l_i+1},
\label{eq:veffective} 
\ee
where the second term on the right-hand side describes the transitions through the intermediate inelastic channels; the real parts of the inelastic loops are absorbed into the low-energy constants $\mathcal{C}_d$ and $\mathcal{C}_f$. 

Then, the Lippmann--Schwinger equation for the $\Upsilon(10860)$ decaying into open-bottom final states can be written as \cite{Wang:2018jlv}
\bea
U_\alpha(M,p)&=&F_\alpha(M,p)\nonumber\\[-1mm]
\label{eq:LSE}\\[-1mm]
&-&\sum_\beta\int
U_\beta(M,q)G_\beta(M,q)V_{\beta\alpha}^{\text{eff}}(M,q,p)\frac{d^3q}{(2\pi)^3},\nonumber
\eea
where $U_\alpha(M,p)$ denotes the physical production amplitude of the $\alpha$-th elastic channel
from a pointlike $S$-wave source, and $F_{B\bar{B}^*}(M,p)=-F_{B^*\bar{B}^*}(M,p)=1$ as dictated by HQSS. 
The Green's function for a two-heavy-meson intermediate state reads
\be\label{Eq:Green}
G_\alpha(M,q)=\frac{2\mu_\alpha}{q^2-p_\alpha^2-i0},\quad p_\alpha^2\equiv 2\mu_\alpha(M-m_{\rm th}^\alpha),
\ee 
where $m_{\rm th}^\alpha$ stands for the $\alpha$-th elastic threshold and $\mu_{\alpha}$ is the reduced mass in this channel. Other components of the multichannel amplitude responsible 
for production of the inelastic channels in the final state can be obtained 
from $U_\alpha(M,p)$ algebraically, which is a consequence of the omitted direct interactions in the inelastic 
channels. In particular, for the $i$-th inelastic channel in the final state we have
\bea
&&\hspace*{-0.03\textwidth}U_i(M,k_i)\nonumber\\[-2mm]
\label{eq:inelastic}\\[-2mm]
&&=-\sum_\alpha\int\frac{d^3q}{(2\pi)^3}U_\alpha(M,q) G_\alpha(M,q)v_{\alpha i}(M,q,k_i),\nonumber
\eea
where the momentum $k_i$ is defined in Eq.~(\ref{eq:mominelastic}) above. It has to be noticed that the Born amplitudes $F_i(M,p)$ coming from the inelastic sources were neglected in Eq.~(\ref{eq:inelastic}). This is justified for the $\pi h_b(mP)$ channels, where the data are dominated by the $Z_b(10610)$ and $Z_b(10650)$ poles emerging from the $B^{(*)}\bar B^*$ dynamics. The corresponding line shapes were included into the combined fit performed in Ref.~\cite{Wang:2018jlv}. On the contrary, in the heavy-spin-conserving $\pi\Upsilon(nS)$ channels, the Born term needs to be kept and the $\pi\pi$ interaction in the final state has to be included. 
How this can be done in a model-independent way will be discussed in detail below.
 
The one-dimensional distributions for the differential widths in the elastic ($B^{(*)}\bar{B}^*$) and inelastic ($\pi h_b(mP)$) channels used in Ref.~\cite{Wang:2018jlv} read
\begin{eqnarray}
\frac{d\Gamma_\alpha}{dM}&=&\frac13\ \frac{2m_{B^{(*)}}2m_{B^*}2m_{\Upsilon(10860)}}{32\pi^3 m_{\Upsilon(10860)}^2}\; p_\pi^* p_\alpha |U_\alpha|^2,\nonumber\\[-2mm]
\\[-2mm]
\frac{d\Gamma_i}{dM}&=&\frac13\ \frac{2m_{h_i}2m_{H_i}2m_{\Upsilon(10860)}}{32\pi^3 m_{\Upsilon(10860)}^2}\; p_\pi^* p_i|U_i|^2,\nonumber
\end{eqnarray}
respectively, where $p_\pi^*$ is the three-momentum of the spectator pion in the rest frame of the $\pi\Upsilon(10860)$ and $p_{\alpha}(p_i)$ is the 
three-momentum in the $\alpha$-th elastic ($i$-th inelastic) channel in the rest frame of the $B^*\bar{B}^{(*)}$ ($\pi\Upsilon(nS)/\pi h_b(mP)$) system. 
Then, the total branching fraction in an elastic or inelastic channel $x$ is defined as 
\be
\Br_x=\frac{\Gamma_x}{\sum_{\alpha=1}^{2}\Gamma_\alpha+\sum_{i=1}^{5}\Gamma_i},
\label{Brx}
\ee
where
\be
\Gamma_x=\int_{M_{\rm min}}^{M_{\rm max}}\left(\frac{d\Gamma_x}{dM}\right)dM,
\ee
and the integral in $M$ covers the entire kinematically allowed region for the considered channel $x$.

The line shapes obtained in the fitting scheme described above are presented in Fig.~\ref{fig:fita}, and the parameters of the fit are listed in Table~\ref{tab:par}. As was explained above, only the total branchings of the $\pi\Upsilon$ channels were included in the fits.

In what follows, the left-hand cut structure of the multichannel production amplitude $U_i$ ($i= \pi \Upsilon(nS)$) from Eq.~\eqref{eq:inelastic}, 
obtained in the framework of the contact Scheme A, will be used as input for a dispersive reconstruction of the $\pi\pi$ FSI.

\section{Final state interaction}
\label{sec:fsi}

\subsection{Kinematics of the reaction}
\label{subsec:kinem}

In this subsection we introduce the kinematics of the decay 
$\Upsilon(10860)(p_i)\to \Upsilon(nS)(p_f)\pi^+(p_1)\pi^-(p_2)$ with $n=1,2,3$. Following a standard approach to such reactions, we built the amplitude $M(s,t,u)$ in a crossed channel, $\Upsilon(p_i)+\Upsilon'(p_f)\to\pi(p_1)+\pi(p_2)$, and define the Mandelstam invariants accordingly,
\be
s=(p_i+p_f)^2,\quad t=(p_f+p_1)^2,\quad u=(p_f+p_2)^2,
\ee
with
\be
p_i^2=m_i^2,\quad p_f^2=m_f^2,\quad p_1^2=p_2^2=m_\pi^2,
\ee
where $m_\pi$, $m_i$, and $m_f$ are the masses of the pion,
$\Upsilon(10860)\equiv\Upsilon$, and $\Upsilon(nS)\equiv\Upsilon'$, respectively. Thus,
\be
s+t+u=m_i^2+m_f^2+2m_\pi^2.
\ee 

In order to proceed, we resort to the kinematics in the centre-of-mass frame of the two pions in the final state, so that ($z \equiv \cos \theta$, where $\theta$ is the angle between the 3-momenta ${\bm p}_1$ and ${\bm p}_f$),
\bea
t(s,z)&=&\frac12(m_i^2+m_f^2+2m_{\pi}^2-s)+\frac12k(s)z,
\nonumber \\[-2mm]
\label{tandu}\\[-2mm]
u(s,z)&=&\frac12(m_i^2+m_f^2+2m_{\pi}^2-s)-\frac12k(s)z,
\nonumber
\eea
where
\be
k(s)=\frac{1}{s}\sqrt{\lambda(s,m_i^2,m_f^2)\, \lambda(s,m_{\pi}^2,m_{\pi}^2)},
\ee
with the function $\lambda$ defined in Eq.~(\ref{lambda}) above.
Consequently, $z$ can be expressed in terms of $t$ and $u$ as
\be
z=\dfrac{t-u}{k(s)}.
\label{z}
\ee

Since the production amplitude for the process 
$\Upsilon(10860)\to \pi^+ \pi^-\Upsilon(nS)$ has the form \footnote{This is correct up to HQSS violating terms and the D-wave operators for the $\Upsilon(nS)$ which do not appear from the mechanisms considered here.}
$$
M^{\rm full} = M(s,t,u)\ \bm{\varepsilon}_{\Upsilon(10860)} \cdot \bm{\varepsilon}^*_{\Upsilon(nS)},
$$
the double differential production rate can be written as 
\begin{equation}
\frac{d ^2 \Br}{d s\, d t}={\cal N} \left|M(s,t,u)\right|^2,
\label{cross-section}
\end{equation}
where the overall normalisation constant $\cal N$ will be fitted to the data. 

\subsection{Dispersive approach to the $\pi\pi$-$K\bar K$ FSI}
\label{sec:disp}

In this subsection we introduce the meson-meson interaction in the final state.
The partial wave decomposition of the amplitude $M(s, t, u)$ reads 
\be
 M(s, t, u)= \sum_l M_l(s) P_l (z),
\ee
where $P_l (z)$ are the Legendre polynomials and the sum runs over all relevant angular momenta $l$. 
We start from the amplitude $M(s,t,u)$ projected onto the $\pi\pi$ $S$-wave,
\be
M_0(s)=\frac12\int_{-1}^{+1}dz M(s,t,u),
\label{Ms}
\ee
which can be split into two pieces,
\be
M_0=M_0^R+M_0^L,
\ee
where the first and second term contain the right- and left-hand cuts only, respectively. The right-hand cut of the amplitude $M_0^R$ comes from the 
FSI while the left-hand cuts of the amplitude $M_0^L$ are due to the dynamics related to the $Z_b$ states. If the contribution $M_0^L$ is known and only the $\pi\pi$ channel is considered for the FSI, the full amplitude can be reconstructed dispersively via 
the solution of the inhomogeneous Omn\`{e}s problem as (see Ref.~\cite{Kang:2013jaa} for a related discussion)
\be
M_0(s)=M_0^L(s)+\frac{\Omega_0(s)}{\pi}\int_{4m_\pi^2}^\infty ds'\frac{M_0^L(s')\sin\delta(s')}{|\Omega_0(s')|(s'-s-i0)},
\label{solMs}
\ee
where $\Omega_0(s)$ is the $S$-wave single-channel \Omnes function\footnote{We use the standard notation $\Omega_l^I$ for the \Omnes function where $l$ and $I$ stand for the partial wave and isospin, respectively. However, since in this work we deal only with isoscalars, the superscript $I=0$ is omitted everywhere.} and $\delta$ is the $\pi\pi$ $S$-wave phase shift (see Appendix~\ref{app:Omega} for details). 
However, given that the energy in the $\pi\pi$ system in the reaction $\Upsilon(10860)\to \pi\pi\Upsilon(1S)$ extends to 1.4 GeV, that is far beyond the $K\bar K$ threshold, the inclusion of the $K\bar K$ component becomes necessary. 
Generalisation of Eq.~(\ref{solMs}) to multiple channels is straightforward,
\bea
&&\hat{M}_0(s)=\hat{M}_0^L(s)\nonumber\\[-1mm]
\label{Mmultch}\\[-1mm]
&&\hspace*{0.02\textwidth}+\frac{\hat{\Omega}_0(s)}{\pi}\int_{4m_\pi^2}^\infty\, ds'\frac{\hat{\Omega}_0^{-1}(s')\hat{T}(s')\hat{\sigma}(s')\hat{M}_0^L(s')}{s'-s-i0}.\nonumber
\eea
Here, the multichannel \Omnes matrix obeys the matrix equation
\be
\hat{\Omega}_0(s)=\frac{1}{\pi}\int_{4m_\pi^2}^\infty ds'\frac{\hat{T}^*(s')\hat{\sigma}(s)\hat{\Omega}_0(s')}{s'-s-i0},
\label{Eq:Omnes}
\ee
where hats indicate multicomponent objects (vectors and matrices), $\hat{\sigma}(s)=\text{diag}\{\sigma_\pi, \sigma_K\}$ is a diagonal matrix with 
$\sigma_P(s)=\sqrt{1-s_P^{\rm th}/s}$, and $s_P^{\rm th}$ for the threshold in the corresponding channel ($P=\pi,K$). In particular, we have $M_0^L=\left([M_0^L]_{\pi\pi},[M_0^L]_{KK}\right)^T$. Furthermore, the $S$-wave meson-meson coupled-channel amplitude $\hat T$ can be parametrised by the the $\pi\pi$ scattering phase shift
 $\delta(s)$ \cite{GarciaMartin:2011cn,Caprini:2011ky,Buettiker:2003pp,Dai:2014zta} as well as the absolute value and phase of the $\pi\pi \to K\bar K$ transition \cite{Buettiker:2003pp,Dai:2014zta}, $g(s)$ and $\psi(s)$, respectively, as 
\bea
&&T(s)=
 \begin{pmatrix}
 T_{\pi\pi\to \pi\pi} & T_{\pi\pi\to K\bar K} \\
 T_{K\bar K\to \pi\pi} & T_{K\bar K\to K\bar K}
 \end{pmatrix}\nonumber\\[-2mm]
 \label{Eq:T}\\[-2mm]
&&\hspace*{0.1\textwidth}=\begin{pmatrix}
 \ds\frac{\eta e^{2i \delta}-1}{2 i \sigma_\pi} & g e^{i\psi} \\
 g e^{i\psi} & \ds\frac{\eta e^{2i (\psi- \delta)}-1}{2 i \sigma_K} 
 \end{pmatrix},\nonumber
 \eea
 where the inelasticity $\eta$ is related to $g$ as
 \be\label{etag}
 \eta =\sqrt{1-4 g^2\, \sigma_{\pi}\,\sigma_{K}\, \theta(s- 4 m_K^2)}.
 \ee

To get the two-pion FSI amplitude one has to consider the component $[\hat{M}_0(s)]_{\pi\pi}$ of the vector (\ref{Mmultch}).
If the amplitude contains contributions from higher partial waves while the FSI is taken into account only in the $S$ wave, one can write
\bea
&&\hat{M}(s,t,u)=\hat{M}_{\text{no-FSI}}(t,u)\nonumber\\[-1mm] 
\label{Mhigher}\\[-1mm]
&&\hspace*{0.02\textwidth}+\frac{\hat{\Omega}_0(s)}{\pi}\int_{4m_\pi^2}^\infty\, ds'\frac{\hat{\Omega}_0^{-1}(s')\hat{T}(s')\hat{\sigma}(s')\hat{M}_0^L(s')}{s'-s-i0},\nonumber
\eea
where $\hat{M}_{\text{no-FSI}}=\hat{M}_0^L+\hat{M}_{\rm higher} $ is the complete tree level production amplitude in the $t$- and $u$-channel, not projected onto partial waves, while the effect of the FSI is taken 
into account by the second term in Eq.~\eqref{Mhigher}. In this study, the $\pi\pi$ component of the production amplitude $\hat{M}_{\text{no-FSI}}$ and its $S$-wave projection $\hat{M}_0^L$ are adopted from 
Ref.~\cite{Wang:2018jlv} --- see Sec.~\ref{sec:LHC} for a detailed discussion. Meanwhile, the resonance production in the channel $\Upsilon(10860) \to K\bar K\Upsilon(nS) $ which proceeds through the $B$- and $B_s$-meson loops is not considered since no information about the $SU(3)$ partners of the $Z_b$ states is available yet. 

The dispersive integral in Eq.~(\ref{Mhigher}), 
\be
\hat I_0(s)\equiv \frac{1}{\pi}\int_{4m_\pi^2}^\infty\, ds'\frac{\hat{\Omega}_0^{-1}(s')\hat{T}(s')\hat{\sigma}(s')\hat{M}_0^L(s')}{s'-s-i0},
\label{Idef}
\ee
where the lower index indicates $l=0$ for the $S$ wave, may need to be subtracted $n$ times to improve convergence and to diminish the role played by the large-$s$ region where the $\pi\pi$ scattering phase is not known well enough. Then, one arrives at
\bea 
&&\hat{I}_0^{(n)}(s)=\hat{\cal P}_{n-1}(s)\nonumber\\[-1mm]
\label{Isub}\\[-1mm]
&&\hspace*{0.1\textwidth}+\frac{s^n}{\pi}\int_{4m_\pi^2}^\infty\, \frac{ds'}{s^{\prime n}}\frac{\hat{\Omega}_0^{-1}(s')\hat{T}(s')\hat{\sigma}(s')\hat{M}_0^L(s')}{s'-s-i0},\nonumber
\eea
where $ \hat{\cal P}_{n-1}(s)$ is a polynomial of the order $n-1$. If the amplitude $\hat{M}_0^L(s)$ has both real and imaginary parts, then the polynomial coefficients are complex numbers. Meanwhile, if there are good reasons to believe that the imaginary part of the amplitude $\mbox{Im}M_0^L(s)$ is controlled by well understood physics (see also a related discussion in Sec.~\ref{sec:match} below), then the imaginary part of the polynomial ${\cal P}_{n-1}(s)$ can be evaluated exploiting sum rules via
\bea\nonumber
\mbox{Im}\hat{\cal P}_{n-1}(s)&=&\sum_{k=0}^{n-1}\frac{s^k}{\pi}\int_{4m_\pi^2}^\infty \frac{ds'}{s^{\prime (k+1)}}\\[-2mm]
\\[-2mm]
 &\times& {\hat{\Omega}_0^{-1}(s')\hat{T}(s')\hat{\sigma}(s')\, \mbox{Im}\hat{M}_0^L(s')},\nonumber
\eea
where it was used that the quantity $\hat{\Omega}_0^{-1}(s')\hat{T}(s')\hat{\sigma}(s')$ is real. 
This allows one to re-write Eq.~(\ref{Isub}) in the form
\bea
&&\hat I_0^{(n)}(s)=\hat R_{n-1}(s)\nonumber\\
&&\hspace*{0.05\textwidth}+\frac{s^n}{\pi}\int_{4m_\pi^2}^\infty\, \frac{ds'}{s^{\prime n}}\frac{\hat{\Omega}_0^{-1}(s')\hat{T}(s')\hat{\sigma}(s')\mbox{Re}\hat{M}_0^L(s')}{s'-s-i0}
\label{Mtot3}\\
&&\hspace*{0.08\textwidth}+\frac{i}{\pi}\int_{4m_\pi^2}^\infty\, {ds'}\frac{\hat{\Omega}_0^{-1}(s')\hat{T}(s')\hat{\sigma}(s')\mbox{Im}\hat{M}_0^L(s')}{s'-s-i0},\nonumber
\eea
where the polynomial $R_{n-1}(s)$ is real by construction, 
\be
\hat R_{n-1}(s)=\mbox{Re}\hat{\cal P}_{n-1}(s),
\ee
and so are its coefficients. As discussed below, Im$\hat{M}_0^L(s)$ is non-vanishing on a finite interval of $s$
only and, accordingly, the integral in the last line of Eq.~(\ref{Mtot3}) does not require any subtractions.

\subsection{The left-hand cut production amplitude }
\label{sec:LHC}

\begin{figure*}[t]
 \begin{center}
 \includegraphics[width=0.45\textwidth]{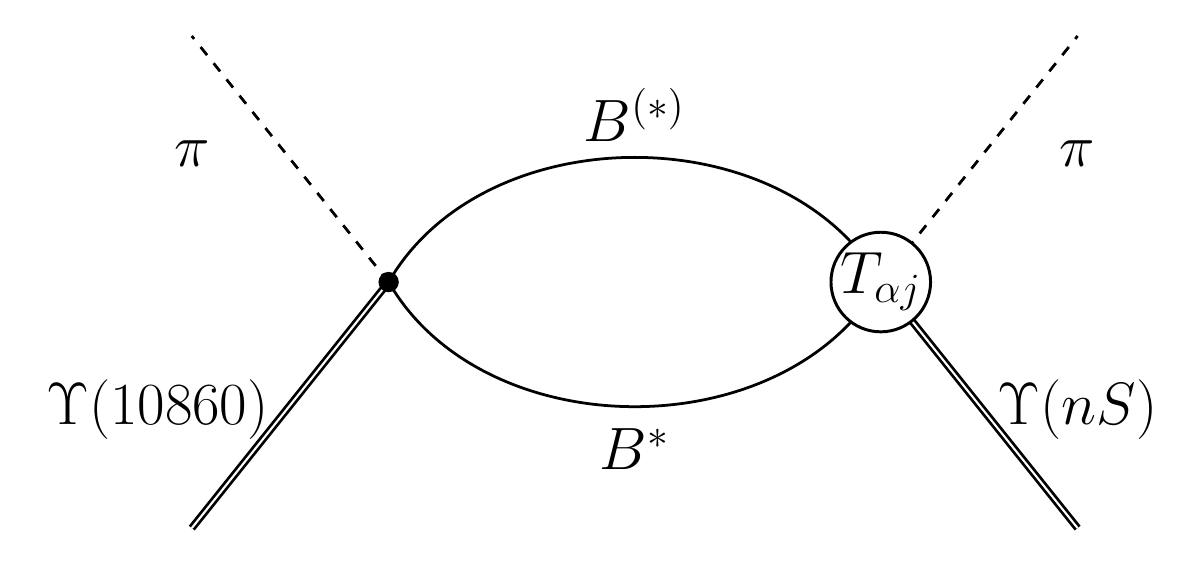}\hspace*{0.05\textwidth}
 \includegraphics[width=0.45\textwidth]{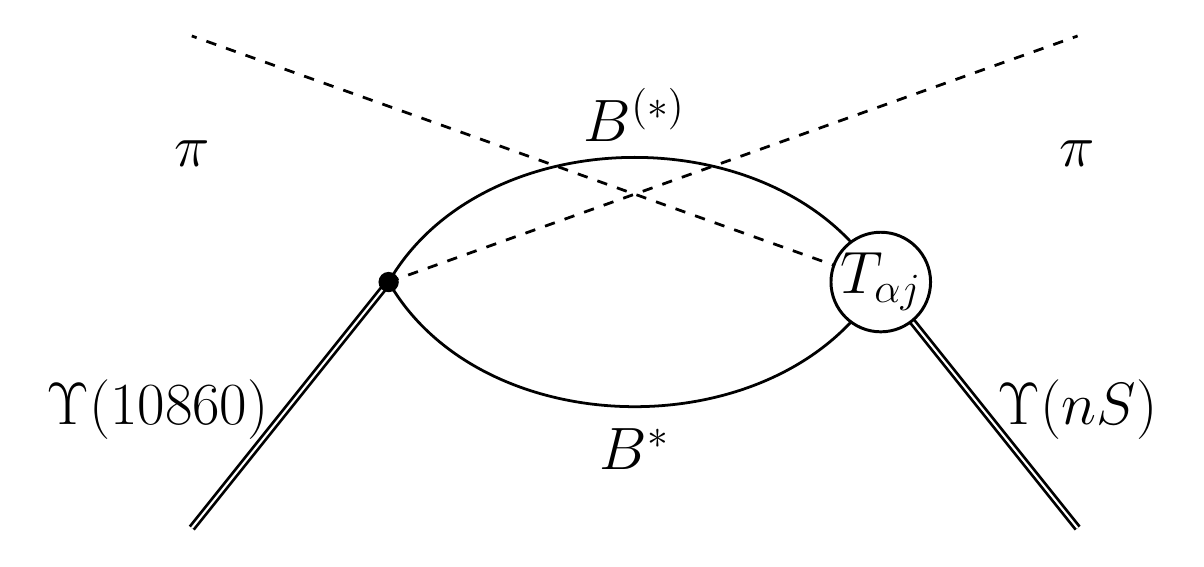} 
 \end{center}
 \caption{Left-hand cuts in the production amplitude $U$ coming from the $B\bar B^*$ and $B^*\bar B^*$ scattering in the $t$- (left) and $u$-channel (right). $T_{\alpha j}$ denotes the coupled-channel amplitude for the transitions 
 $B^{(*)}\bar B^*\to \pi\Upsilon(nS)$.}
 \label{fig:LHC}
 \end{figure*}

In order to proceed with the formulae derived in the previous subsection, we need to specify the form of the production amplitude $M_{\text{no-FSI}}$ introduced in Eq.~(\ref{Mhigher}) and determine its
$S$-wave projection,
\be
M_0^L(s)=\frac12\int_{-1}^1dz \, M_{\text{no-FSI}}(t,u).
\label{MBorns}
\ee

Consider, as a preliminary step, a stable-$Z_b$ exchange in the $t$- and $u$-channel. Then, assuming pointlike $\Upsilon^{(\prime)}\to\pi Z_b$ vertices, 
up to an overall constant, one can write the invariant Born amplitude as 
\be
M_{\rm stable}(t,u;m_z)=\frac{1}{t-m_z^2} +\frac{1}{u-m_z^2},
\label{MBorn0}
\ee
where $m_z$ is the mass of the mentioned stable $Z_b$ particle. For future convenience we specify this mass as an argument of the amplitude.

Performing the partial wave projection as introduced in Eq.~(\ref{MBorns}) for the Born amplitude (\ref{MBorn0}) and using the prescription $m_i^2\to m_i^2+i0$ 
one arrives at 
\begin{widetext}
\be
M_{0,\text{stable}}^{L}(s,m_z)=
\left\{
\begin{array}{ll}
\ds -\frac{2}{\kappa(s)}\left(\log\frac{Y(s)+\kappa(s)}{Y(s)-\kappa(s)}
+2\pi i\,\theta(s_a-s)\theta(s-s_+)\right),& (m_f + m_{\pi})^2 < m_z^2 < \frac12(m_f^2+m_i^2)-m_\pi^2,\\[4mm]
\ds -\frac{2}{\kappa(s)}\left(\log\frac{Y(s)+\kappa(s)}{Y(s)-\kappa(s)}
+2\pi i\,\theta(s_a-s)\right) 
, & m_z^2 < (m_f + m_{\pi})^2 < \frac12(m_f^2+m_i^2)-m_\pi^2.
\end{array}
\right.
\label{MsLC0full}
\ee
\end{widetext}
Here
\bea
&&Y(s)=s+2m_z^2-m_i^2-m_f^2-2m_\pi^2,\nonumber\\[-2mm]
\label{Eq:logdef}\\[-2mm]
&&\kappa(s)=\sigma_\pi(s)\sqrt{\lambda(s,m_i^2,m_f^2)},\nonumber
\eea
and 
\be
s_a=2m_\pi^2+m_f^2+m_i^2-2m_z^2
\label{sa}
\ee 
is the root of the equation $Y(s)=0$. Furthermore, the logarithmic branch points $s_\pm$ (also known as anomalous thresholds) found as the roots of the equation $t (s_\pm,z=\pm 1)=m_z^2$ read
\bea
s_\pm &=& \frac{(m_i^2-m_f^2)^2}{4m_z^2}\nonumber\\[-1mm] 
\label{s_min_pl}\\[-1mm]
&-& \frac{\left(\sqrt{\lambda({m_i^2,m_{\pi}^2,m_z^2)}}\pm \sqrt{\lambda({m_f^2,m_{\pi}^2,m_z^2}})\right)^2 }{4m_z^2},\nonumber
\eea
where $\lambda$ is the triangle function from Eq.~\eqref{lambda}.

 In the regime 
\be
(m_f + m_{\pi})^2 < m_z^2< \frac12(m_f^2+m_i^2)-m_\pi^2,
\ee
$s_+$ is real and the anomalous threshold generates only a phase term which is included in the first formula in Eq.~\eqref{MsLC0full} (see also Ref.~\cite{Molnar:2019uos} for a related discussion). However, for 
\be
m_z^2 < (m_f + m_{\pi})^2 < \frac12(m_f^2+m_i^2)-m_\pi^2,
\ee 
the branch point $s_+$ becomes complex and the dispersive integral defined in Eq.~\eqref{Mtot3} acquires an additional anomalous contribution calling for an integration along some complex path (see Appendix~\ref{app:anom} for details). 
Namely, using $M_{0;{\rm anom}}^{L} =-4\pi i/\kappa$ for the anomalous discontinuity, the integral $\hat I_0^{(n)}(s)$ from Eq.~\eqref{Mtot3} gets modified as 
$$
\hat I_0^{(n)}(s) \to \hat I_0^{(n)}(s) + \hat I_0^\text{anom}(s,m_z),
$$
where
\be
\hat I_0^\text{anom}(s, m_z) =
\frac{s^n}{2\pi i}\int_{0}^1\, \frac{dx}{\zeta^n} \frac{d\zeta}{dx} \frac{8\pi}{\kappa(\zeta)}\frac{\hat{\Omega}_0^{-1}(\zeta)\hat{T}(\zeta)\hat{\sigma}(\zeta)}{\zeta-s-i0},
\label{Eq:anom}
\ee
and $\zeta=(1-x) s_+ + x\, 4 m_{\pi}^2$ is the straight-line path between the two-pion threshold and the branch point
of the logarithm, $s_+$. 

A crucial point of the coupled-channel approach developed in Ref.~\cite{Wang:2018jlv} is that the resonances $Z_b$ are not introduced as asymptotic states of the theory but appear as near-threshold poles of the amplitude fitted to the data. This implies that, instead of the stable $Z_b$ propagator used in Eq.~(\ref{MBorn0}), the inelastic amplitude $U_i$ ($i=\pi\Upsilon(nS)$ with $n=1,2,3$) from Ref~\cite{Wang:2018jlv}, generated through the $B$-meson loops and
evaluated as given in Eq.~(\ref{eq:inelastic}), provides the input for building 
$M_{\text{no-FSI}}$ and $M_0^L$ --- see Fig.~\ref{fig:LHC} for its diagrammatic representation. 
To proceed, we employ a dispersive representation for the production amplitude $U$ (to simplify notations we omit the inelastic index $i$
and thus consider a particular inelastic final state),
\bea
&&M_{\text{no-FSI}}(t,u)=U(t)+U(u)\nonumber\\ 
&=&-\frac{1}{\pi}\int_{\mu^2_{\rm min}}^{\mu^2_{\rm max}} d\mu^2\, \mbox{Im\,}U(\mu^2) 
\left(\frac{1}{t-\mu^2} + \frac{1}{u-\mu^2}\right)\label{Eq:Mprod} \\ 
&=&
\int_{\mu^2_{\rm min}}^{\mu^2_{\rm max}} d\mu^2 {\rho(\mu^2)} M_{\rm stable}(t,u;\mu),\nonumber
\eea
 where we used Eq.~\eqref{MBorn0} and introduced the spectral function 
\be
\rho(\mu^2)=-\frac{1}{\pi}\mbox{Im\,}U(\mu^2).
\label{rhoFitA}
\ee
The lower limit in the integral above is given by the lowest relevant threshold which may contribute to the imaginary part of the amplitude. Unitarity 
of the production amplitudes $U(\mu^2)$ requires integration from 
the lowest inelastic threshold $\pi\Upsilon(1S)$, although the leading contributions start from the $B\bar B^* \pi$ threshold.
The upper limit of integration in Eq.~\eqref{Eq:Mprod} should formally be infinite, however the use of a finite momentum regulator in the Lippmann-Schwinger equations restricts the maximal values of the on-shell momenta which therefore cannot exceed the cutoff $\Lambda$. Thus in practical calculations $\mu_{\rm max}=m_{\rm th}^\alpha + \Lambda^2/2\mu_\alpha$ is used \cite{Wang:2018jlv} (see also Eq.~(\ref{Eq:Green}) for relevant definitions), and we have verified that the dispersive representation from Eq.~\eqref{Eq:Mprod} reproduces $U(t)+U(u)$ from Ref.~\cite{Wang:2018jlv} quite precisely.
The limit of a stable particle, Eq.~(\ref{MBorn0}), is reached from Eq.~\eqref{Eq:Mprod} for $\rho(\mu^2)=\delta(\mu^2-m_z^2)$.
With the mass distribution of the $Z_b$ states included, the $S$-wave partial wave amplitude reads 
\be
M_0^L(s)=\int_{\mu^2_{\rm min}}^{\mu^2_{\rm max}} d\mu^2 \rho(\mu^2) M_{0, \text{stable}}^{L}(s,\mu),
\label{Eq:M0}
\ee
where $M_{0, \text{stable}}^{L}(s,\mu)$ is defined in Eq.~\eqref{MsLC0full}.
Also, the anomalous term from Eq.~\eqref{Eq:anom} has to be weighted with the spectral function to read
\be
\hat I_0^\text{anom}(s) =\int_{\mu^2_{\rm min}}^{(m_{\pi}+m_f)^2} d\mu^2 \rho(\mu^2) \hat I_0^\text{anom}(s, \mu).
\label{Eq:anom2}
\ee

\subsection{Matching to chiral perturbation theory}
\label{sec:match}

\begin{figure*}[t]
 \begin{center}
 \begin{tabular}{cccc}
 \epsfig{file=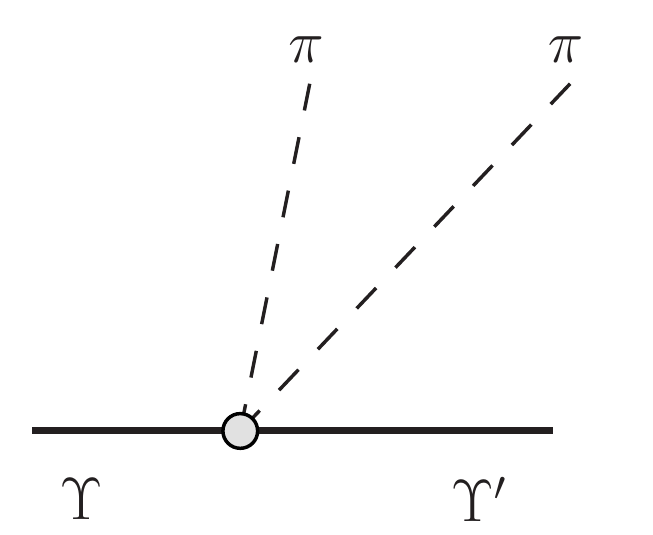,width=0.23\textwidth}&
 \epsfig{file=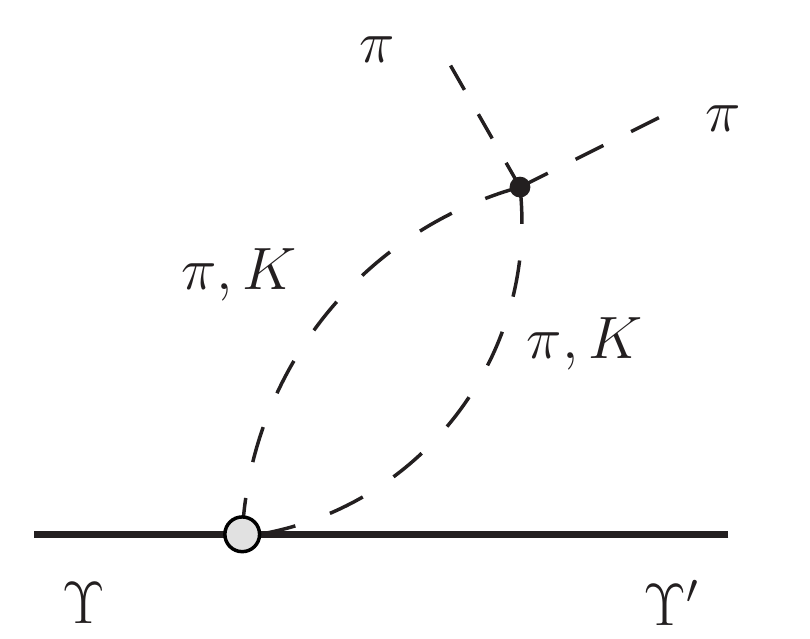,width=0.26\textwidth}&
 \raisebox{-2.5mm}{\epsfig{file=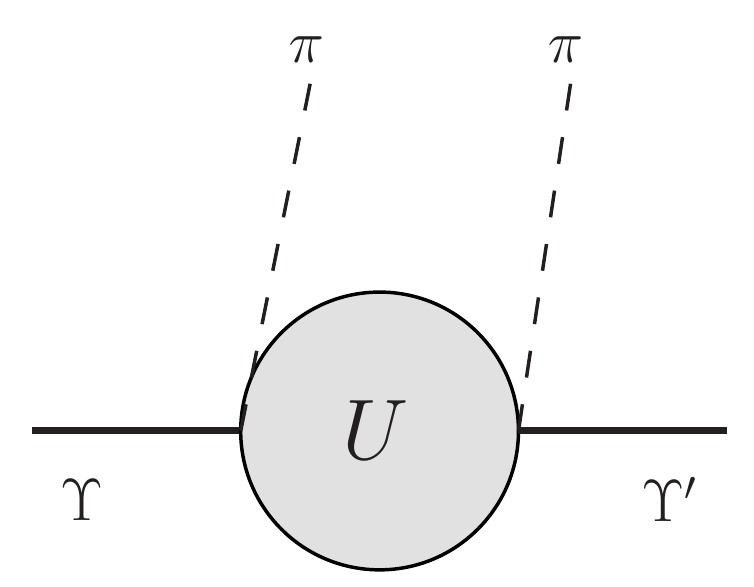,width=0.23\textwidth}}&
 \raisebox{-2.5mm}{\epsfig{file=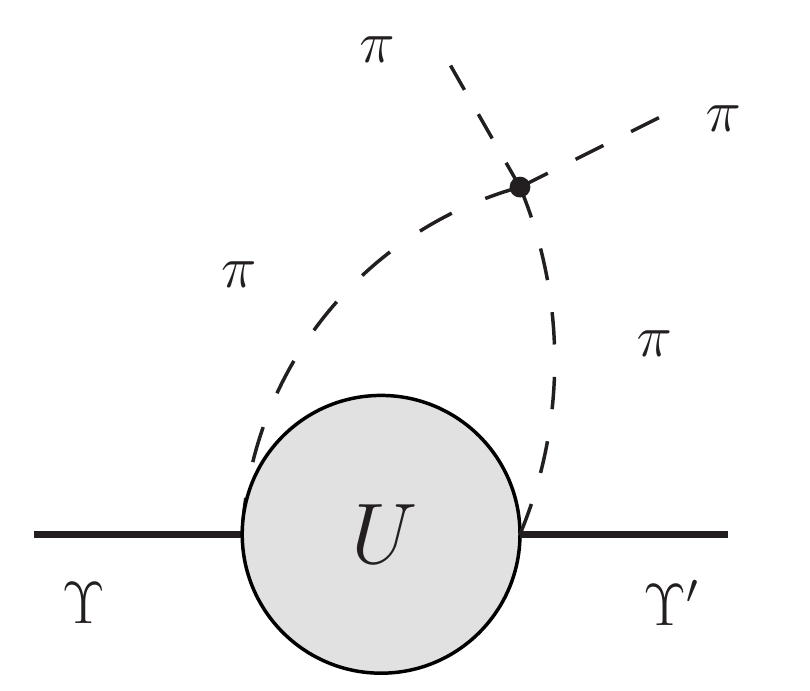,width=0.25\textwidth}}\\[2mm]
 (a)&(b)&(c)&(d)
 \end{tabular}
 \end{center}
 \caption{Diagrams contributing to the full amplitude $M(s,t,u)$ from Eq.~(\ref{Mfull}) for the decay $\Upsilon\to\pi\pi\Upsilon'$ ($\Upsilon\equiv\Upsilon(10860)$, $\Upsilon'\equiv\Upsilon(nS)$ with $n=1,2,3$): (a) the contact diagram; (b) the contact diagram with the $\pi\pi$ and $K\bar K$ FSI; (c)  the production amplitude $M_{\text{no-FSI}}$ in the $t$- and $u$-channel which contains left-hand cuts from the $Z_b$'s generated in a coupled-channel approach 
 of Ref.~\cite{Wang:2018jlv}, see also Eq.~\eqref{Eq:Mprod}; (d) same as in (c) but with the $\pi\pi$ FSI.}
 \label{fig:contact}
 \end{figure*}

Two comments on the convergence of the subtracted dispersive integral (\ref{Mtot3}) entering Eq.~(\ref{Mhigher}) are in order here. 
First, as follows from Eqs.~\eqref{MsLC0full} and \eqref{s_min_pl}, 
the imaginary part of the $M_0^L(s)$ may be nonzero only at a finite interval in $s$ between the logarithmic branch points $s_-$ and $s_+$.
 Therefore, the integral from $\mbox{Im}M_0^L$ in the last term in Eq.~(\ref{Mtot3}) is finite. The appearance of an imaginary part in the left-hand cut amplitudes is a specific consequence of the cuts which are allowed in the process 
$ \Upsilon(10860)\to \pi\pi\Upsilon(nS)$, especially from the $B^{(*)}\bar B^*\pi$ intermediate states but also from inelastic channels. While inelastic cuts can also contribute to similar decays from the $\Upsilon(3S)$ and $\Upsilon(4S)$, the presence of the cuts in the elastic channels is only allowed kinematically starting from the $\Upsilon(10860)$. 
 
 The number of subtractions $n$ needed to render the dispersion integral convergent, can be determined by the high-energy behaviour of the function 
 $\hat{\Omega}_0^{-1}(s')\hat{T}(s')\hat{\sigma}(s') \mbox{Re}\hat{M}_0^L(s')$. 
 Since the \Omnes matrix is determined by an unsubtracted 
 dispersion relation \eqref{Eq:Omnes}, each of its elements behaves as $1/s$ at high energies --- see Ref.~\cite{Moussallam:1999aq} for details. Thus each element of its inverse scales as $s$.
 For the scattering amplitude $\hat T$ in the two-channel case defined in Eq.~\eqref{Eq:T} it is possible to demonstrate that for $T_{12} \propto 1/s^{3/2}$ at large $s$~\cite{Hoferichter:2012wf}, 
 $T_{11}$ and $T_{22} $ need to scale as $1/s^3$. Therefore, we proceed with a conservative estimate that $\hat T(s) \propto 1/s^{3/2}$ at large $s$. 
 The left-hand cut amplitude for a stable particle $M_{0, \text{stable}}^{L}(s,m_z)$ falls off as $ \log(s)/s$ at large $s$, however the full left-hand cut production amplitude $\hat{M}_0^L$ may decrease slower and is expected to approach a constant. Then, even without subtractions, one in principle should arrive at a convergent integral. However, 
to suppress the contribution of the large-$s$ region, where the details of the $\pi\pi$ interaction are badly known, 
in what follows, twice subtracted dispersive integrals are considered, and the polynomial in Eq.~(\ref{Mtot3}) takes the form
\be
R_1(s)=a+bs,
\label{R1ab}
\ee
with real parameters $a$ and $b$, as was explained above. In what follows, it will be shown that one of these constants is mostly redundant at least for production of $\Upsilon(nS)$ with $n=2$ and 3. 

It is important to notice that the polynomial $R_1(s)$ parametrises the amplitude for $\Upsilon\Upsilon'\pi\pi$
at small values of $s$ and as such can be matched
to chiral perturbation theory. Specifically, in the limit of switching off the final-state interactions, $\delta(s) \to 0$, $g(s) \to 0$ in Eq.~\eqref{Eq:T} and thus setting
$\Omega_0(s)\to 1$, the subtraction
functions must agree with the chiral amplitudes corresponding to the direct transitions $\Upsilon\to \pi\pi\Upsilon'$ \cite{Chen:2015jgl}. 

If one introduces spin multiplets for heavy-heavy fields, 
\bea
J &=& \vec{\Upsilon} \cdot \boldsymbol{\sigma}+\eta_b,\nonumber
\eea
then the effective Lagrangian for the contact
$\Upsilon\Upsilon^{\prime}\pi\pi$ and
$\Upsilon\Upsilon^{\prime}K\bar{K}$ coupling, at the lowest order in the chiral and heavy-quark expansions,
reads \cite{Mannel:1995jt,Chen:2015jgl,Chen:2016mjn}
\begin{equation}\label{LagrangianUpUppipi}
{\cal L}_{\Upsilon\Upsilon^{\prime}\Phi\Phi} = \frac{c_1}{2}\bra J^\dagger
J^\prime \ket \bra u_\mu u^\mu\ket +\frac{c_2}{2}\bra J^\dagger
J^\prime \ket \bra u_\mu u_\nu\ket v^\mu v^\nu +\mathrm{h.c.},
\end{equation}
where $v^\mu$ is the 4-velocity of the heavy quark. The contribution of the pseudoscalar Goldstone bosons for the spontaneous breaking of the chiral symmetry can be parametrised as
\bea
&u_\mu=i\left(u^\dagger\partial_\mu u-u\partial_\mu u^\dagger\right),&\nonumber \\
&\ds u = \exp \left( \frac{i\Phi}{\sqrt{2}f}\right),& \\
&\ds \Phi = \begin{pmatrix}
 {\frac{1}{\sqrt{2}}\pi ^0 +\frac{1}{\sqrt{6}}\eta _8 } & {\pi^+ } & {K^+ } \\
 {\pi^- } & {-\frac{1}{\sqrt{2}}\pi ^0 +\frac{1}{\sqrt{6}}\eta _8} & {K^0 } \\
 { K^-} & {\bar{K}^0 } & {-\frac{2}{\sqrt{6}}\eta_8 } \\
 \end{pmatrix},&\nonumber 
 \label{eq:u-phi-def}
\eea
where $f$ is the pseudo-Goldstone boson decay constant, $f_\pi=92.2$ MeV and $f_K=113.0$ MeV. If one makes an expansion in the (soft) pion momenta $q_\pi$, both operators quoted in Eq.~\eqref{LagrangianUpUppipi} scale as $\mathcal{O}(q_\pi^2)$ .

Considering an $S$-wave contribution for the tree-level amplitudes, $\hat M_0^{\chi}(s) =\left(M_0^{\chi, \pi\pi}(s), \frac2{\sqrt{3}}M_0^{\chi,KK}(s) \right)^T $, one finds ($P=\pi,K$)
\bea\label{M0chi}
&& M_0^{\chi,PP}(s)=-\frac{2}{f_P^2}\sqrt{m_\Upsilon m_{\Upsilon'}}\bigg\{c_1 \left(s-2m_P^2 \right)\nonumber\\[-2mm]
\label{eq.M0chiral}\\[-2mm]
&&\hspace*{0.17\textwidth}+\frac{c_2}{2} \bigg[s+q^2\left(1 -\frac{\sigma_P^2(s)}{3}
\right)\bigg]\bigg\},\nonumber 
\eea
where $q$ is the 3-momentum of the final $\Upsilon'$ in the
rest frame of the initial $\Upsilon$, that is,
\be \label{eq:q}
q=\frac{1}{2m_\Upsilon}
\lambda^{1/2}\left(m_\Upsilon^2,m_{\Upsilon'}^2,s\right).
\ee
Up to some small corrections, the amplitude \eqref{M0chi} behaves as a linear polynomial in $s$. 
Thus the chiral amplitude at low energies depends on the two low-energy constants (LECs) $c_1$ and $c_2$ which can be treated as fitting parameters instead of $a$ and $b$ from Eq.~(\ref{R1ab}). This amplitude corresponds to the contact diagram depicted in Fig.~\ref{fig:contact}(a).

Then, the amplitude $M(s,t,u)$ from Eq.~\eqref{cross-section}, which now includes the effects from the $\pi\pi$ and $K\bar K$ FSI in the $S$-wave, takes the form
\be
M(s,t,u)=M_{\text{no-FSI}}(t,u)+\hat\Omega_0(s) \left(\hat M_0^{\chi,\pi\pi}(s)+\hat{\tilde{I}}_0^{(2)}(s)\right),
\label{Mfull}
\ee
where $M_{\text{no-FSI}}$ is given in Eq.~(\ref{Eq:Mprod}) and the $\pi\pi$ component \{11\} of the matrix multiplication is implied. The integral $\hat{\tilde{I}}_0^{(2)}$ is defined as
(\emph{cf.} Eq.~(\ref{Mtot3}))
\bea\nonumber
\hat{\tilde I}_0^{(2)}(s)&=& 
\frac{s^2}{\pi}\int_{4m_\pi^2}^\infty\, \frac{ds'}{s^{\prime 2}}\frac{\hat{\Omega}_0^{-1}(s')\hat{T}(s')\hat{\sigma}(s')\, \mbox{Re}{M}_0^L(s')}{s'-s-i0}
\nonumber\\
&+&\frac{i}{\pi}\int_{4m_\pi^2}^\infty\, {ds'}\frac{\hat{\Omega}_0^{-1}(s')\hat{T}(s')\hat{\sigma}(s')\, \mbox{Im}{M}_0^L(s')}{s'-s-i0}\nonumber\\
&+& \hat I_0^\text{anom}(s),
\label{ttildedef}
\eea
where ${M}_0^L(s')$ and $\hat I_0^\text{anom}(s)$ are given in Eqs.~\eqref{Eq:M0} and \eqref{Eq:anom2}, respectively. 
 
The diagrams representing different contributions to the amplitude of Eq.~(\ref{Mfull}) are depicted in Fig.~\ref{fig:contact}: the sum of the diagrams (a) and (b) corresponds to the term 
$ \hat\Omega_0(s) \hat M_0^{\chi,\pi\pi}(s)$, the diagram (c) gives the production amplitude $M_{\text{no-FSI}}$ in the $t$- and $u$-channel, and the diagram (d) describes the contribution of the last term $\hat\Omega_0(s)\hat{\tilde{I}}_0^{(2)}(s)$. 

\subsection{Inclusion of the $\pi\pi$ FSI in the $D$ wave}
\label{sec:Dwave}

Generalisation of Eq.~(\ref{Mfull}) to the $\pi\pi$ FSI in higher partial waves is straightforward,
\bea
&&M(s,t,u)=M_{\text{no-FSI}}(t,u)\nonumber\\[-2mm]
\label{Mfullgen}\\[-2mm]
&&\hspace*{0.1\textwidth}+\sum_l\hat\Omega_l(s) \left(\hat M_l^{\chi,\pi\pi}(s)+\hat{\tilde{I}}_l^{(n_l)}(s)\right),\nonumber
\eea
where the sum runs over all relevant angular momenta $l$.
More specifically, taking into account the $\pi\pi$ interaction in the $D$ wave, we write for the amplitude
\bea
&&
M(s,t,u)=M_{\text{no-FSI}}(t,u)+\hat\Omega_0(s) \left(\hat M_0^{\chi,\pi\pi}(s)+\hat{\tilde{I}}_0^{(2)}(s)\right)\nonumber\\
&&\hspace*{0.24\textwidth}+\Omega_2(s)M_2^{\chi,\pi\pi}(s)P_2(z),
\label{MfullD}
\eea
where $P_2(z)$ is the second-order Legendre polynomial (see also Eq.~(\ref{z})), the amplitude $M_2^{\chi,\pi\pi}(s)$ extracted from the Lagrangian (\ref{LagrangianUpUppipi}) reads
\be
M_2^{\chi,\pi\pi}(s)=\frac{2}{3f_\pi^2}\sqrt{m_\Upsilon m_{\Upsilon'}}c_2q^2\sigma_\pi^2(s),
\label{M2chi}
\ee
and the diagrams which correspond to the amplitude (\ref{M2chi}) coincide with those depicted in Fig.~\ref{fig:contact} (a) and (b), however with no kaons in the loop. No additional parameter is involved in 
 the amplitude (\ref{M2chi}), since $c_2$ also enters Eq.~\eqref{M0chi}. 
The $D$-wave \Omnes function $\Omega_2(s)$ in Eq.~\eqref{M2chi} is calculated using the $D$-wave $\pi\pi$ phase shift from Ref.~\cite{GarciaMartin:2011cn} and is dominated by the $f_2(1270)$ resonance contribution. 
In general, the amplitude in Eq.~\eqref{MfullD} should also contain the dispersive integral $\tilde{I}_2^{(n_2)}(s)$, which is however neglected in the current study. 
This is motivated by the fact that the corresponding $D$-wave contribution from the chiral polynomial in Eq.\eqref{M2chi} plays only a very minor role in the fits, as discussed in Sec. \ref{sec:analysis}. 
While in this study the main focus is put on the development of the appropriate formalism and testing the general consistency of the coupled-channel EFT approach of Ref.~\cite{Wang:2018jlv} 
with the data in the $\Upsilon(10860)\to \pi\pi \Upsilon(nS)$ decays, we postpone the calculation of $\tilde{I}_2^{(n_2)}(s)$ to a future global analysis of all data available in various channels. 

\section{Data analysis}
\label{sec:analysis}

With the coupled-channel approach developed in Ref.~\cite{Wang:2018jlv} and further augmented by the $\pi\pi/K\bar K$ interaction in the final state, as explained in the previous section, we are in a position to analyse the data on the decays $\Upsilon(10860)\to\pi\pi\Upsilon(nS)$ ($n=1,2,3$). We use the Belle data from Ref.~\cite{Garmash:2014dhx} and make a maximum likelihood ${\cal L}$
fit to the two-dimensional distributions. The minimised function is defined in a standard way \cite{Zyla:2020zbs},
\be
-2\log {\cal L}=2\sum_i \left( \mu_i-n_i+n_i\log\frac{n_i}{ \mu_i}\right),
\ee
where the sum runs over all bins in the analysed two-dimensional distribution, while $n_i$ and $ \mu_i$ are the number of events and the value of the theoretical signal function in the $i$-th bin, respectively.
The signal function is corrected for the efficiency, and the 
experimental background distribution is added. To account for the invariant mass resolution in $M^2(\pi\Upsilon(nS))$, we convolute the theoretical results with the Gaussian resolution function with 
$\sigma = 4.3$ MeV, 2.2 MeV and 1.3 MeV for $\Upsilon(1S)$, $\Upsilon(2S)$ and $\Upsilon(3S)$, respectively. The effect of the energy resolution for the invariant mass $M^2(\pi\pi)$ is neglected, since the line shapes for this projection 
do not show any sharp peaking structures and are quite smooth functions of $M^2(\pi\pi)$.
In each $\pi\pi\Upsilon(nS)$ ($n=1,2,3$) channel we fit three parameters: $c_1$, $c_2$, and ${\cal N}$, where the former parameters correspond to the low-energy constants in the chiral polynomial [see Eq.~(\ref{eq.M0chiral})] while the latter one provides the overall normalisation of the distribution [see Eq.~(\ref{cross-section})]. We exclude the region near
the Dalitz plot boundary to minimise the effects of the $e^+e^-$ centre-of-mass
energy spread and the detector resolution. 

 \begin{table} 
 \begin{tabular}{|c|c|c|c|}
 \hline 
 Channel & $c_1\times 10^4$ [GeV$^{-1}]$& $c_2\times 10^4$ [GeV$^{-1}]$& Correlation\\
 \hline
 $\pi\pi\Upsilon(1S)$ &$0.322 \pm 0.017$ &$-0.171\pm 0.020 $& 61\%\\
 $\pi\pi\Upsilon(2S)$ & $21.1 \pm 0.6$ & $-12.6 \pm 0.5$ & 97\%\\
 $\pi\pi\Upsilon(3S)$ & $-17.8 \pm 3.1$& $16.1 \pm 3.6 $& 91\%\\
 \hline
 \end{tabular}
 \caption{Parameters of the fits to the two-dimensional Dalitz plots obtained in this work. The last column shows the correlation between the parameters.} \label{tab:fitparams}
 \end{table}

\begin{figure*}[t]
\begin{center}
\epsfig{file=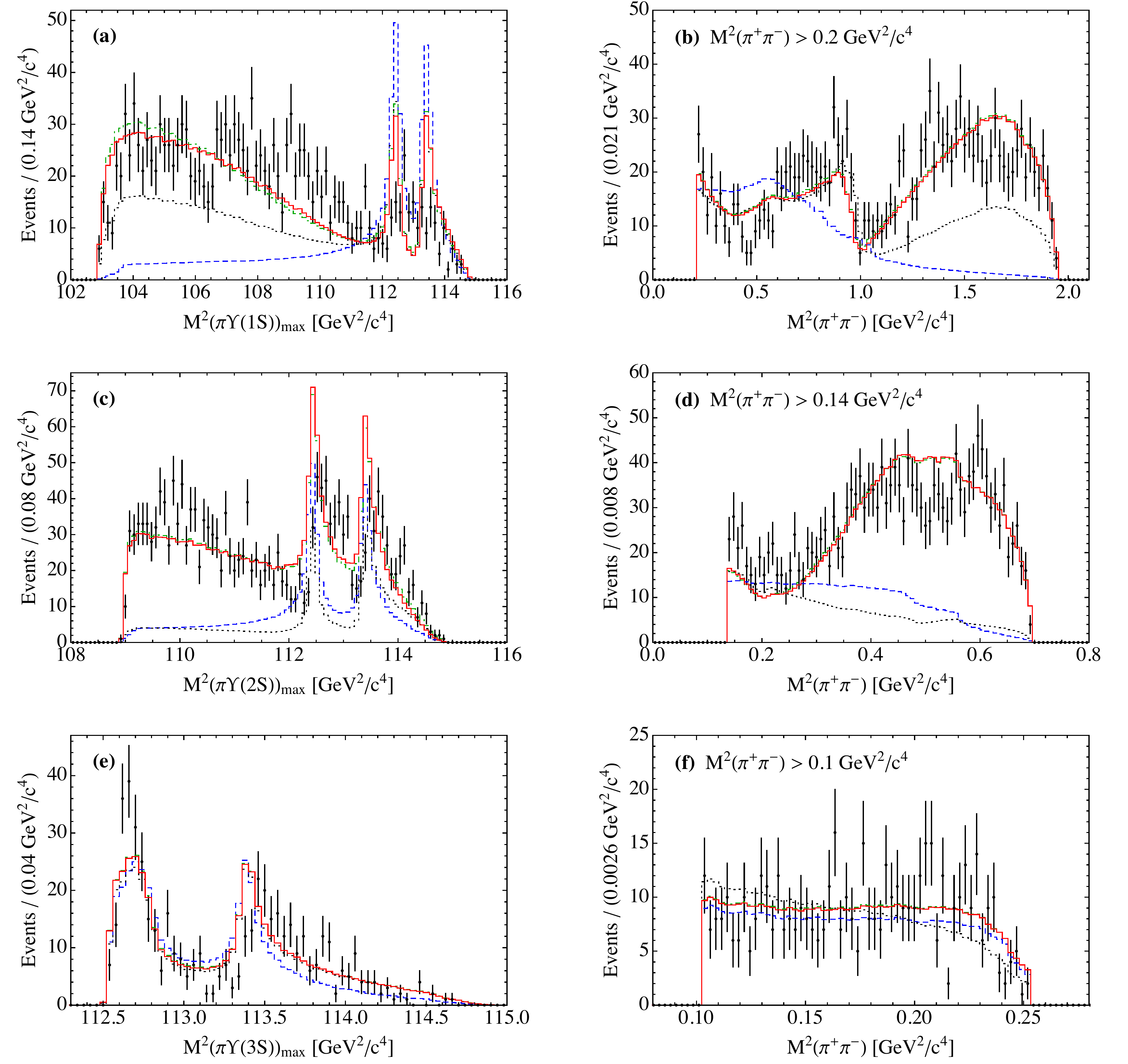,width=1.\textwidth}
\end{center}
\caption{One-dimensional projections on the $M^2(\pi\Upsilon(nS))$ ($n=1,2,3$) (left panel) and $M^2(\pi\pi)$ (right panel) of the fits to the two-dimensional Dalitz plots, as explained in the text. 
The blue dashed lines correspond to the results from the $t$- and $u$-channel contributions from Ref.~\cite{Wang:2018jlv} plus experimental background (the latter is however negligible for $\Upsilon(2S)$ and $\Upsilon(3S)$), the black dotted lines correspond to the same result plus the dispersively reconstructed $s$-channel contribution of the amplitudes from Ref.~\cite{Wang:2018jlv} (the integral $\tilde{I}_0$) while the green lines include in addition the effect from chiral polynomials. The red solid lines represent the full result, where the difference from the green lines is due to the $D$-wave contributions. The theoretical results are convoluted with the energy resolution in $M^2(\pi\Upsilon(nS))$, see text. 
The experimental data are shown as black data points with uncertainties.}
\label{fig:fitsUps}
\end{figure*}

The results of the data analysis performed in this work are presented in Fig.~\ref{fig:fitsUps} in the form of one-dimensional projections of the Dalitz plots, namely the invariant mass distributions $M^2(\pi\pi)$ and $M^2(\pi\Upsilon(nS))$. The parameters of the fits are listed in Table~\ref{tab:fitparams}, where the last column shows that at least for production of $\Upsilon(2S)$ and $\Upsilon(3S)$ only a linear combination of the two constants is relevant. 
From Fig.~\ref{fig:fitsUps} one can see that the developed approach is able to describe the data in the $\pi\pi\Upsilon$ channels rather well. In each plot we give several curves showing different contributions to the total rate. The relative importance of these contributions changes significantly with the mass of the bottomonium $\Upsilon(nS)$ in the final state.
\begin{itemize}
\item In case of the $\pi\pi\Upsilon(3S)$ channel, given a very limited phase space available, the distributions are dominated by the $t$- and $u$-channel production amplitudes from Ref.~\cite{Wang:2018jlv} ($M_{\text{no-FSI}}$ in 
Eq.~\eqref{Mhigher}) which capture the gross features of the experimental distributions. Effects from the $\pi\pi$ FSI and chiral polynomials are marginal individually and, in addition, numerically cancel each other to a large extent. 
\item For the $\pi\pi\Upsilon(2S)$ channel, the resulting contribution from the $t$- and $u$-channel amplitudes and the dispersively reconstructed $s$-channel FSI are all important but not sufficient
to explain the data. The missing strength and the energy dependence comes from the chiral contact terms with the $\pi\pi$ FSI. 
\item Finally, in case of the $\pi\pi\Upsilon(1S)$ final state with the maximal available phase space, 
the pattern of the individual contributions is qualitatively similar to that in the $ \Upsilon(2S)$ channel. In particular, the $t$- and $u$-channel production amplitudes provide the peaking structures from the $Z_b$'s and enhance the 
 small $\pi\pi$ region in the corresponding line shape. The dispersive integral $I_0$, enhanced by the $\pi\pi$-$K\bar K$ FSI from the coupled-channel \Omnes function, is very important. It provides more than 
 half of the full signal at small $M_{\pi \Upsilon(nS)}^2$ and large $M_{\pi \pi}^2$ but also significantly
 affects the $M_{\pi \Upsilon(nS)}^2$ line shape in the $Z_b$'s area. In addition, $I_0$ drives the shape of the $M_{\pi \pi}^2$ spectrum near the $K\bar K$ 
 threshold and together with the $t$- and $u$-channel production amplitudes describes the low $M_{\pi \pi}^2$ region. The residual contribution comes from chiral contact terms in $S$ wave, while their effect in $D$ wave is minor. 
\end{itemize}

\begin{figure*}[t]
\begin{center}
\epsfig{file=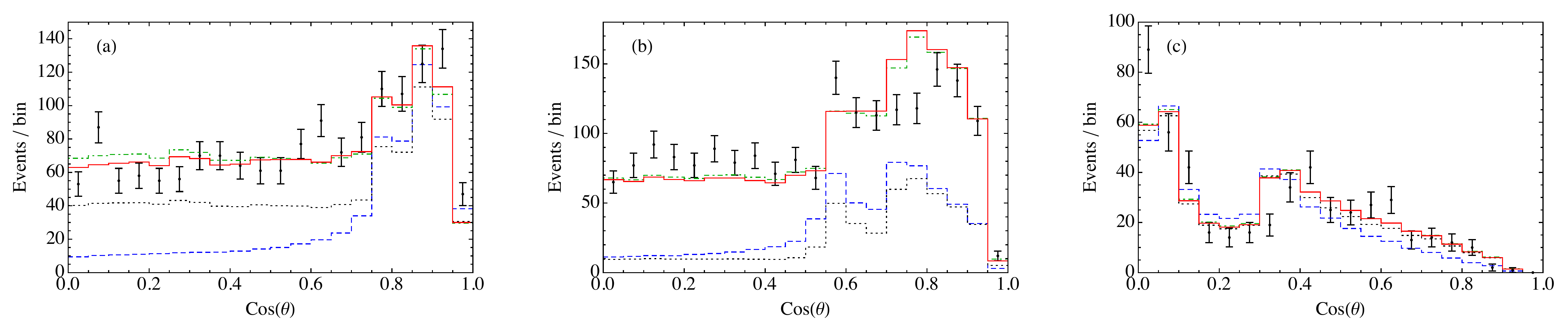,width=1.\textwidth}
\end{center}
\caption{Helicity angular distributions for the decays $\Upsilon(10860)\to \pi\pi\Upsilon(nS)$ with $n=1,2,3$ for the left, middle, and right plot, respectively. The helicity angle $z=\cos \theta$ is defined in Eq.~\eqref{z}. 
For the lines definition see the caption of Fig.~\ref{fig:fitsUps}.}
\label{fig:angle}
\end{figure*}

In Fig.~\ref{fig:angle} we give the helicity angular distributions, including individual contributions to it. The
anisotropies of these distributions are largely driven by the higher partial-wave contributions from the amplitude $M_{\text{no-FSI}}(t,u)$ in Eq.~\eqref{MfullD} --- as usual the partial wave expansion in some subsystem converges badly, as soon as there is a narrow, near on-shell state in the crossed channel. 
 
While the approach advocated in this study allows for a quite reasonable quantitative understanding of the line shapes in the $\pi\pi \Upsilon(nS)$ channels, we would like to emphasise 
that the peaks of the $Z_b$'s in our results (red solid curves in Fig.~\ref{fig:fitsUps}), which by construction are consistent with the experimental $B^{(*)}\bar{B}^*$ and 
$\pi h_b(mP)$ ($m=1,2$) distributions, are not exactly in accord with the data in the $\pi \Upsilon(nS)$ channels. 
This observation still calls for an explanation.

\section{Conclusions}
\label{sec:conslusions}

In this work we extended the coupled-channel approach developed previously in Ref.~\cite{Wang:2018jlv} to incorporate the $\pi\pi$ FSI in the heavy-quark-conserving channels $\pi\pi\Upsilon(nS)$ ($n=1,2,3$). 
Maximum likelihood fits to the two-dimensional Dalitz plots are performed in each such channel. 
Compatibility with the previous analysis of the line shapes in the $B^{(*)}\bar{B}^*$ and $\pi h_b(mP)$ ($m=1,2$) channels is guaranteed by employing the amplitudes obtained in Ref.~\cite{Wang:2018jlv} as input for the current research. 

The $\pi\pi$ FSI was incorporated into the coupled-channel scheme employing a dispersive approach, in which the left-hand cut contributions were provided by the previously found inelastic production amplitudes. 
 Also, it was conjectured that the dominating sources of the imaginary parts of the production amplitudes in the $\pi\pi\Upsilon(nS)$ channels are fully under control and, therefore, only the real parts of the complex polynomials 
 were fitted to the data, while their imaginary parts appeared as actual predictions of the approach.
As a cross-check, it was verified that the fits did not improve much if the imaginary parts of the chiral polynomials were also fitted to the data. 

The results obtained demonstrate that the extended approach developed in this work is able to describe the existing experimental data with a reasonable accuracy. As expected, the role of the FSI
increases with the increase of the allowed phase space. For example, it is remarkable, that the results for the $\pi\pi\Upsilon(3S)$ channel, which are almost completely driven by the production operators from Ref.~\cite{Wang:2018jlv} with all the parameters fixed from other data, are in a relatively good agreement 
with the data for the $\pi\pi$ and $\pi \Upsilon(3S)$ projections. 
Meanwhile, effects from the $\pi\pi$ FSI in the $S$-wave are found to be important for the $\pi\pi\Upsilon(2S)$ and, especially, $\pi\pi\Upsilon(1S)$ channel. 
In the latter case, also the $K\bar K$ channel 
plays a very important role. In particular, the coupled-channel $\pi\pi$-$K\bar K$ dynamics shows up as a very clear dip structure near the $K\bar K$ threshold, which is consistent with the data.
The effect of the $D$-wave resonance $f_2(1270)$ included via the $D$-wave \Omnes function together with the contact production operators is found to be very minor for all channels. 

Among the open questions to mention is an observation that the peaks corresponding to the $Z_b$'s from the presented analysis do not completely correspond to the data in the $\pi \Upsilon(nS)$ channels.
To address this question a full combined fit to all measured production and decay channels for the $Z_b$'s should be performed. This lies beyond the scope of the present research. 

We consider the results presented here as an important step towards a comprehensive combined analysis of the data in all production and decay channels of the $Z_b$'s, which should in the future also incorporate pion exchanges. Such an analysis which will include the full information contained in the two-dimensional distributions for the $\pi\pi\Upsilon(nS)$ channels should provide the most accurate determination of the parameters of the theory
and, as a result, allow making quite precise predictions for the line shapes, pole positions, decays couplings and other properties of the yet unobserved spin partner states $W_{bJ}$. Such an insight is expected to provide an important theoretical background for the searches of the $W_{bJ}$'s as well as other near-threshold exotic candidates in the experiment Belle II and, possibly, in LHCb and future hadronic experiments.

\begin{acknowledgments}
We are grateful to A. E. Bondar for valuable discussions.
This work is supported in part by the National Natural Science Foundation of China (NSFC) under Grants No.~11835015, No.~11961141012 and No.~11947302, by the NSFC and the Deutsche Forschungsgemeinschaft (DFG) through the funds provided to the Sino-German Collaborative Research Center ``Symmetries and the Emergence of Structure in QCD'' (NSFC Grant No. 11621131001, DFG Grant No. CRC110), by the Chinese Academy of Sciences (CAS) under Grants No.~XDB34030303 and No.~QYZDB-SSW-SYS013, and by the CAS Center for Excellence in Particle Physics (CCEPP). 
Work of V.B., R.M., and A.N. was supported by the Russian Science Foundation (Grant No. 18-12-00226). 
\end{acknowledgments}

\appendix

\section{Dispersive approach to the amplitude $M(s,t,u)$}
\label{app:Omega}

Under the assumption that the right-hand cuts of the amplitude (\ref{Ms}) are only due to the FSI of pion pairs, one can write for the discontinuity of $M_0$ on this cut
\bea
\mbox{Disc\,}M_0(s)&=&M_0(s+i\epsilon)-M_0(s-i\epsilon)
\equiv M_{s+}-M_{s-}\nonumber\\[-1mm]
\\[-1mm]
&=&2 i \, T^*(s)\sigma_\pi(s)M_0(s)\theta(s-4m_\pi^2),\nonumber
\eea
where $s$ is the $\pi\pi$ invariant energy, $\sigma_\pi(s)=\sqrt{1-4m_\pi^2/s}$, and $T(s)=\sin\delta e^{i \delta}/\sigma_\pi(s)$ is the $\pi\pi$ scattering amplitude. Then the standard way to proceed is to introduce the \Omnes function $\Omega_0(s)$ which by definition obeys the equation
\bea\label{Eq:discOmega}
\mbox{Disc\,}\Omega_0(s)&=&2 i \, T^*(s)\sigma_\pi(s)\Omega_0(s)\theta(s-4m_\pi^2)\nonumber\\[-2mm]
\\[-2mm] 
&=& 2i \sin\delta|\Omega_0|\theta(s-4m_\pi^2),\nonumber
\eea
so that
\be
\mbox{Disc\,}M_0=\frac{\mbox{Disc\,}\Omega_0(s)}{\Omega_0(s)}M_0.
\label{D1}
\ee
Since $\mbox{Disc\,}M_0^L=0$ on the unitary cut and, therefore, $\mbox{Disc\,}M_0=\mbox{Disc\,}M_0^R$, this can be re-written as
\be
\mbox{Disc\,}M_0^R=\frac{\mbox{Disc\,}\Omega_0(s)}{\Omega_0(s)}(M_0^R+M_0^L).
\label{D2}
\ee

The solution of Eq.~(\ref{D2}) for $M_0^R$ can be found in the form
\be
M_0^R=\Omega_0(s)G(s),
\label{M1}
\ee
where $G(s)$ is an unknown function to be restored dispersively from its discontinuity. To this end we find from Eq.~(\ref{M1}) that
\bea
\mbox{Disc\,}M_0^R&=&\mbox{Disc\,}(\Omega_0 G)=\Omega_{0+}G_+-\Omega_{0-}G_-
\nonumber\\
&=&\Omega_{0+}G_+-\Omega_{0-}G_++\Omega_{0-}G_+-\Omega_{0-}G_-\nonumber\\[-2mm]
\\[-2mm]
&=&G_+\mbox{Disc\,}\Omega_0+\Omega_{0-}\mbox{Disc\,}G\nonumber\\
&=&\frac{\mbox{Disc\,}\Omega_0}{\Omega_0} M_0^R+|\Omega_0|e^{-i\delta}\mbox{Disc\,}G,\nonumber
\label{D3}
\eea
where $X_\pm\equiv X(s\pm i\epsilon)$ for an arbitrary function $X(s)$, and
it was used that
\be
\Omega_{0+}\equiv\Omega_0=|\Omega_0|e^{i\delta},\quad \Omega_{0-}=\Omega_{0+}e^{-2i\delta}=|\Omega_0|e^{-i\delta}.
\ee

Equations~(\ref{D2}) and (\ref{D3}) together yield
\be
\mbox{Disc\,}G(s)=\frac{\mbox{Disc\,}\Omega_0}{|\Omega_0|^2}M_0^L,
\label{DG}
\ee
and, with the help of Eq.~(\ref{Eq:discOmega}), one finds
\be
\frac{\mbox{Disc\,}G(s)}{2\pi i}=\frac{1}{\pi}\frac{\sin\delta}{|\Omega_0|}M_0^L.
\ee

Then, the function $G(s)$ is restored as
\be
G(s)= \frac{1}{\pi}\int_{4m_\pi^2}^\infty{ds'}\frac{M_0^L(s')\sin\delta(s')}{|\Omega_0(s')|(s'-s-i0)},
\ee
and the solution for the full amplitude takes the form of Eq.~(\ref{solMs}).

If the amplitude $M_0^L(s)$ has no imaginary part, it is easy to verify that the phase of the amplitude $M_0$ is given by the $\pi\pi$ scattering phase $\delta$. Indeed, employing the Sokhotski–Plemelj formula,
\be
\frac{1}{s'-s-i0}=\mbox{p.v.}\frac{1}{s'-s}+i\pi\delta(s'-s),
\label{sokh}
\ee
where p.v. stands for the principal value, one can re-write Eq.~(\ref{solMs}) in the form
\bea
M_0&=&M_0^Le^{i\delta}\cos\delta+
\frac{\Omega_0(s)}{\pi}\fint_{4m_\pi^2}^\infty ds'\frac{M_0^L(s')\sin\delta(s')}{|\Omega_0(s')|(s'-s)}\nonumber\\[-1mm]
\label{Mtot1new}\\[-1mm]
&=&e^{i\delta}\left(M_0^L\cos\delta+
\frac{|\Omega_0(s)|}{\pi}\fint_{4m_\pi^2}^\infty ds'\frac{M_0^L(s')\sin\delta(s')}{|\Omega_0(s')|(s'-s)}\right),\nonumber
\eea
where it was used that
\be
1+ie^{i\delta}\sin\delta=e^{i\delta}\cos\delta.
\ee
Since the expression in parentheses is real by construction, the entire phase of the amplitude $M_0$ is indeed given by the $\pi\pi$ scattering phase $\delta$.

Generalization of the formalism to the coupled-channel case is straightforward and amounts to replacing the scalar objects by matrices, as given in Eq.~\eqref{Mmultch} in Sec.~\ref{subsec:kinem}. 

\section{Anomalous contributions to the amplitude $M_0^L(s)$}
\label{app:anom}

\begin{figure}[t]
 \begin{center}
 \epsfig{file=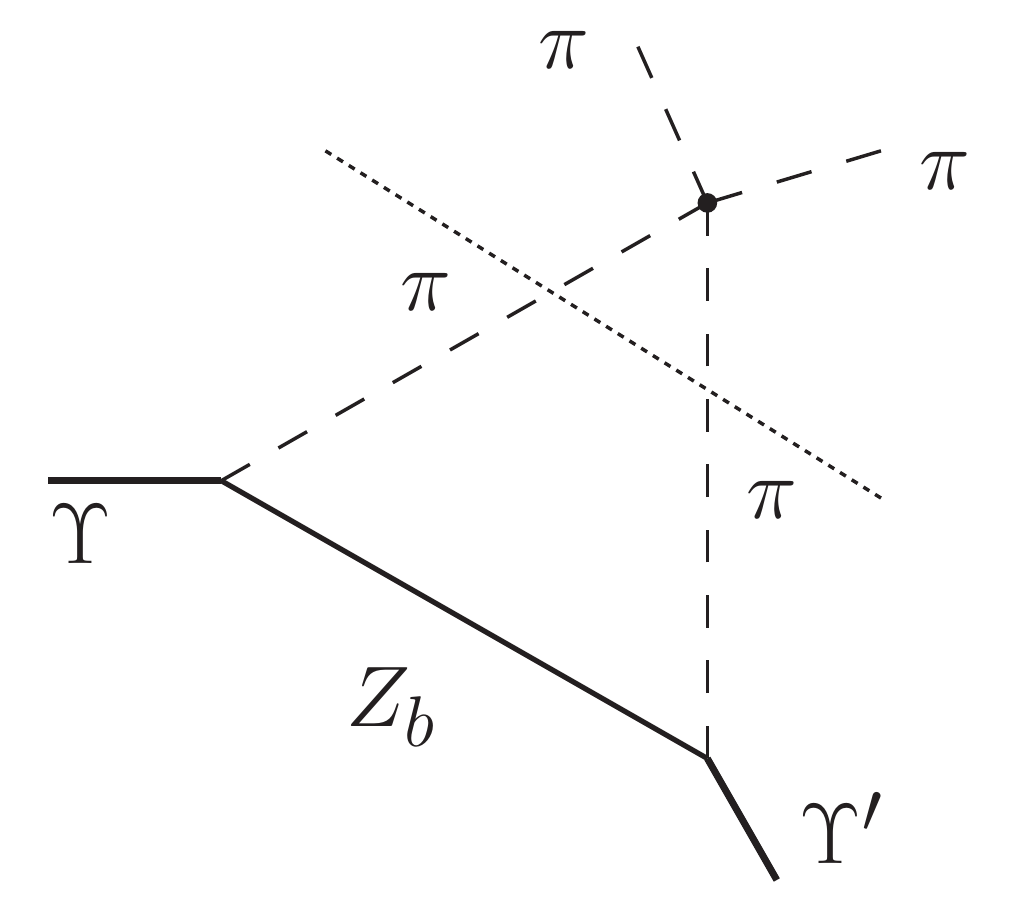,width=0.30\textwidth}
 \end{center}
 \caption{The triangle diagram with the $\pi$-$\pi$ FSI described by the scalar loop function $C_0$ as introduced in Eq.~(\ref{C0def}). The dotted line indicates the cut --- see Eq.~(\ref{C0mznaive}).}\label{fig:pipi}
 \end{figure}

Consider a triangle diagram depicted in Fig.~\ref{fig:pipi} with the $\pi\pi$ interaction in the final state (the $\pi\pi$ vertex is considered pointlike),
\begin{widetext}
\be
C_0(s)=\frac{1}{i\pi^2}\int d^4l\frac{1}{[l^2-m_z^2+i0][(l+p_i)^2-m_\pi^2+i0][(l-p_f)^2-m_\pi^2+i0]},
\label{C0def}
\ee
\end{widetext}
where a shorthand notation is used for the standard scalar loop function $C_0(s)\equiv C_0(m_i^2,s,m_f^2,m_z^2,m_\pi^2,m_\pi^2)$. The function $C_0(s)$ can be calculated through a dispersive integral,
\be
C_0(s)=\frac{1}{2\pi i}\int_{4m_\pi^2}^\infty ds'\,\frac{\mbox{Disc\,}C_0(s')}{s'-s},
\label{C0}
\ee
where the discontinuity $\mbox{Disc\,}C_0(s)$ is simply related to the $S$-wave projected left-hand cut amplitude (\ref{MBorns}) for a stable particle with the mass $m_z$, 
\be
M_{0, \text{stable}}^{L}(s,m_z)=\left(-\frac{2}{\sigma_\pi(s)}\right)\frac{\mbox{Disc\,}C_0(s)}{2\pi i}.
\label{discdef}
\ee
Therefore, studies of the amplitude $M_0^L(s)$ amount to building an analytical form of the discontinuity $\mbox{Disc\,}C_0(s)$ which reproduces the tabulated triangle loop function $C_0(s)$ in the kinematical regimes of interest through the dispersive integral (\ref{C0}). 

Typically, the discontinuity of a loop function is related to an appropriate cut of the corresponding diagram (see the dotted line in Fig.~\ref{fig:pipi})
evaluated with the help of the Cutkosky rules. For a heavy $Z_b$ with the mass $m_z^2 >\frac12(m_f^2+m_i^2)-m_\pi^2$, this gives
\bea
&&\left[\frac{\mbox{disc}\ C_0(s)}{2\pi i}\right]_{\rm naive}=\frac12 \sigma_\pi(s)\int_{-1}^1 \frac{dz}{t-m_z^2}\nonumber\\[-1mm]
\label{C0mznaive}\\[-1mm]
&=&-\frac{\sigma_\pi(s)}{\kappa(s)} \left(\log\frac{Y(s)+\kappa(s)}{Y(s)-\kappa(s)} +2\pi i\theta(s_a-s)
\right),\nonumber
\eea
where the analytic continuation $m_i^2\to m_i^2 +i0$ is implied, $Y(s)$ and $\kappa(s)$ are defined in Eq.~\eqref{Eq:logdef}, and the point $s_a$ in Eq.~(\ref{sa}). The steplike function on the right-hand side of Eq.~(\ref{C0mznaive}) provides an appropriate phase of the logarithm when $Y(s)=0$ and its argument changes the sign. 
The discontinuity (\ref{C0mznaive}) is labelled as naive since, for $m_z^2 < \frac12(m_f^2+m_i^2)-m_\pi^2$, the logarithm branch point $s_+$, defined in Eq.~\eqref{s_min_pl} as the root of the equation $Y(s)+\kappa(s)=0$, 
 appears on the physical Riemann sheet. As a consequence, the discontinuity $\mbox{Disc\,}C_0(s)$ acquires an additional anomalous term which does not correspond any more to simply cutting the triangle diagram in the pion lines --- see, for example, Ref.~\cite{Junker:2019vvy}. Thus, in order to avoid crossing of the logarithm cut spread between $s_-$ and $s_+$ with the unitarity cut, the contour of integration in the dispersive integral needs to be deformed.
This gives
\be
\delta C_0^{\rm anom}(s)=\frac{1}{2\pi i}\int_0^1dx\ \frac{d\zeta(x)}{dx}\frac{\mbox{Disc\,}C_0^{\rm anom}(\zeta(x))}{\zeta(x)-s},
\label{C0anom}
\ee
where
\be
\frac{\mbox{Disc\,}C_0^{\rm anom}(\zeta(x))}{2\pi i}= -2\pi i\frac{\sigma_\pi(\zeta(x))}{\kappa(\zeta(x))}, 
\label{discC0anom}
\ee
and the path of integration $\zeta(x)$ in Eq.~(\ref{C0anom}) can be chosen straight-line,
\be
\zeta(x)=xs_{\rm th}+(1-x)s_+,
\ee
with $s_{\rm th}=4m_\pi^2$ for the two-pion threshold and $s_+$ which appears on the first sheet. 

 \begin{figure}[t]
 \begin{center}
 \epsfig{file=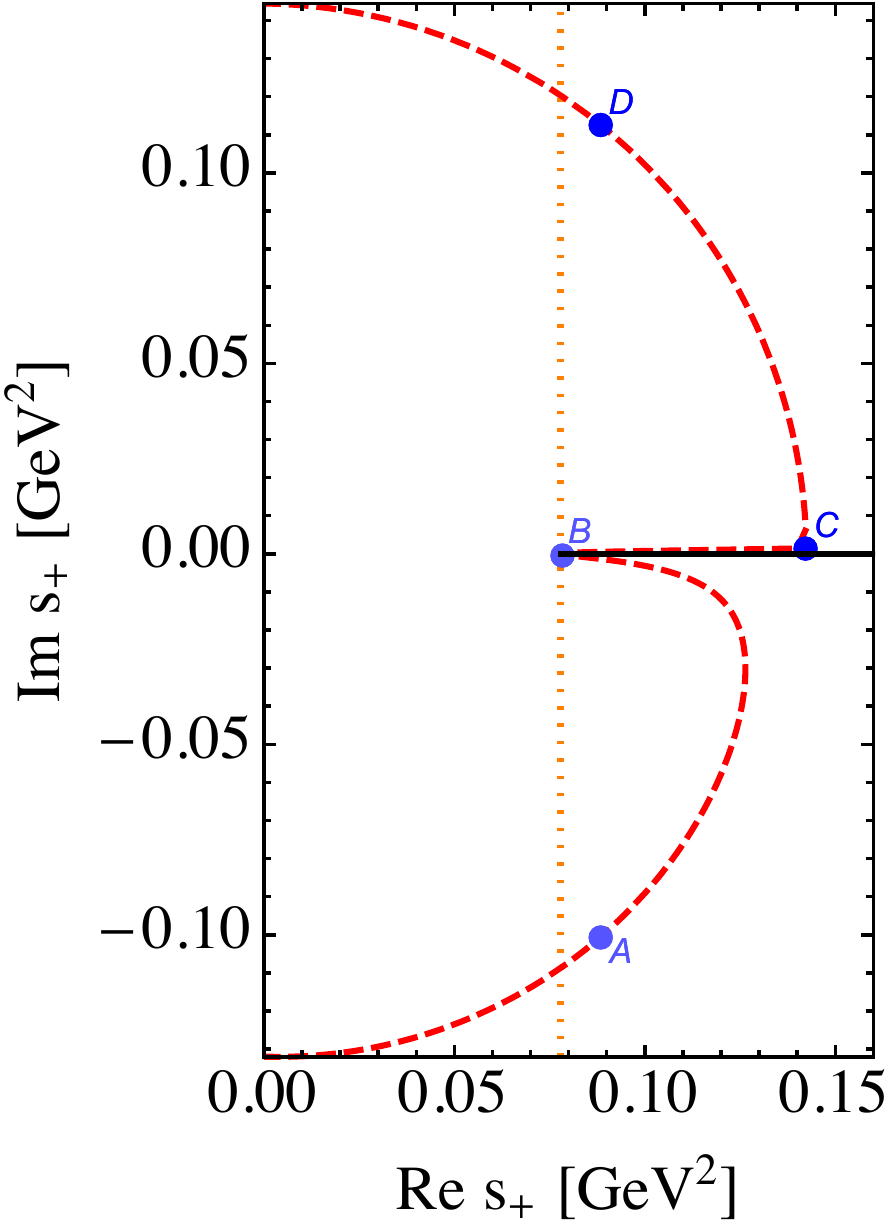,width=0.36\textwidth}
 \end{center}
 \caption{\label{fig:spsm} The motion of the logarithm branch point $s_+$ (the dashed red line) defined in Eq.~(\ref{s_min_pl}) in the $s$ complex plane as the ``running'' mass of the $Z_b$ varies---see the discussion in the text. 
 The convention $m_i^2\to m_i^2+ i \epsilon$ is employed and, to guide the eye, some finite tiny $\epsilon$ is set.
 The unitarity cut is shown by black solid line. It starts at $4 m_{\pi}^2$, as indicated by the vertical dotted orange line. 
The points marked with the capital latin letters, from A to D, are added to illustrate the relevant regions for $s_+$. 
The figure is plotted for the masses of the reaction $\Upsilon(10860)\to \Upsilon(3S)\pi\pi$.}
 \end{figure} 

The interested reader can find a relevant discussion on an anomalous contribution in Ref.~\cite{StefanPhD}.
Here we only briefly consider the kinematic regimes important for the problem at hand, as illustrated in Fig.~\ref{fig:spsm}. Moreover, we only focus on the kinematic range with $s\geqslant s_\text{th}$ which is relevant for our study. 
We start from the case of a heavy $Z_b$ with the mass $m_z^2 >\frac12(m_f^2+m_i^2)-m_\pi^2$, which is also satisfied for the physical masses of both $Z_b(10610)$ and $Z_b(10650)$
quoted by the Particle Data Group~\cite{Zyla:2020zbs}. This case corresponds to the interval between points A and B in Fig.~\ref{fig:spsm}. The discontinuity in this regime is given by Eq.~\eqref{C0mznaive}, and no anomalous term is required. 
Also, since $s_a$ is negative, the $\theta$-function term in Eq.~\eqref{C0mznaive} vanishes. 
For $m_z^2 =\frac12(m_f^2+m_i^2)-m_\pi^2$, the branch point $s_+$ hits the two-pion threshold at the point B and then, for 
$m_z^2 < \frac12(m_f^2+m_i^2)-m_\pi^2$, the anomalous contribution emerges, since $s_+$ enters the first sheet. In the regime $(m_f+m_\pi)^2 <m_z^2 < \frac12(m_f^2+m_i^2)-m_\pi^2$ (see the interval between points B and C in Fig.~\ref{fig:spsm}), 
both $s_+$ and $s_-$ are real and the following relations hold
\be
s_{\rm th}<s_+<s_- <s_a \ll (m_i+m_f)^2.
\ee

\begin{table}[t!] 
    \begin{tabular}{|c|c|c|}
    \hline 
    Channel & $m_f + m_{\pi}$, GeV& $\sqrt{\frac12(m_f^2+m_i^2)-m_\pi^2}$, GeV\\
    \hline
    $\pi\pi\Upsilon(1S)$ &9.5999&10.1833\\
    $\pi\pi\Upsilon(2S)$ &10.1626&10.449\\
    $\pi\pi\Upsilon(3S)$ &10.4948&10.6097\\
    \hline
    \end{tabular}
    \caption{ \label{tab:anom} The critical value of the running mass $m_z$ when $s_+$ emerges in the first sheet but is still real (second column) and when $s_+$ goes to the complex plane (first column). } 
    \end{table}

It is easy to verify that the effect of an anomalous threshold in this case amounts to simply $2\pi i$ terms added to the logarithm between the threshold and the branch point $s_+$.
Then the resulting discontinuity reads (\emph{cf.} the naive formula (\ref{C0mznaive}))
\bea
&&\left[\frac{\mbox{disc}\ C_0(s)}{2\pi i}\right]_{m_i^2+i0}=
-\frac{\sigma_\pi(s)}{\kappa(s)} \Bigl(\log\frac{Y(s)+\kappa(s)}{Y(s)-\kappa(s)}\nonumber\\[-2mm]
\label{C0mz}\\[-2mm]
&&\hspace*{0.15\textwidth}+2\pi i\theta(s-s_+)\theta(s_a-s)\Bigr),\nonumber
\eea
and it is straightforward to verify that it indeed restores the standard loop function $C_0(s)$ through the dispersion relation (\ref{C0}).
On the other hand, if $m_z$ decreases below the value $m_f+m_\pi$ [that is, $ m_z^2< (m_f + m_{\pi})^2 < \frac12(m_f^2+m_i^2)-m_\pi^2$] the branch point $s_+$ goes to the complex plane (see the path between points C to D in Fig.~\ref{fig:spsm})
and the dispersive reconstruction of the integral $C_0$ reads
\bea
\hspace{-1cm}C_0(s)=\frac{1}{2\pi i}\int_{4m_\pi^2}^\infty ds'\,\frac{\mbox{Disc\,}C_0(s')_{\rm naive}}{s'-s} +\delta C_0^{\rm anom}(s),
\label{C0anom2}
\eea
where $\delta C_0^{\rm anom}(s)$ does not reduce to $2\pi i$ any more.

In Table~\ref{tab:anom}, we list the critical values of the running mass $m_z$ when the anomalous term becomes relevant for the $\pi\pi\Upsilon(nS)$ ($n=1,2,3$) channels.

\end{document}